\documentstyle[12pt,epsfig]{article}
\voffset     = - 0.10 in
\begin{document}
\thispagestyle{empty}

\begin{center}
{\Large\bf
Hadron Energy Reconstruction for the ATLAS Barrel Prototype 
Combined Calorimeter in the Framework of the Non-parametrical Method 
}
\end{center}

\bigskip

\begin{center}
{\large\bf 
Y.A.~Kulchitsky$^{a, b,}$\footnote{
E-mail: Iouri.Koultchitski@cern.ch}, 
M.V.~Kuzmin$^{a, b}$, \\
J.A.~Budagov$^{b}$,
V.B.~Vinogradov$^{b}$, \\
M.~Nessi$^{c}$
} 

\bigskip
\bigskip

{\sl $^{a}$ 
Institute of Physics, National Academy of Sciences, Minsk, Belarus}
\\
{\sl $^{b}$ Joint Institute for Nuclear Research, Dubna, Russia}
\\
\smallskip
{\sl $^{c}$ CERN, Geneva, Switzerland}
\end{center}



\vspace*{\fill}

\begin{abstract}
Hadron energy reconstruction for the Atlas barrel prototype combined 
calorimeter, consisting of the lead-liquid argon electromagnetic part and 
the iron-scintillator hadronic part, in the framework of the 
non-parametrical method has been fulfilled.
This me\-thod uses only the known $e/h$ ratios and the electron calibration 
constants and does not require the determination of any parameters by a 
minimization technique. 
The obtained reconstruction of the mean values of energies is within 
$\pm 1\%$ and the fractional energy resolution is 
$[(58\pm3)\% \sqrt{GeV}/\sqrt{E}+(2.5\pm0.3)\%]\oplus (1.7\pm0.2)\ GeV/E$.
The obtained value of the $e/h$ ratio for electromagnetic compartment of the
combined calorimeter is $1.74\pm0.04$ and agrees with the prediction that 
$e/h > 1.7$ for this electromagnetic calorimeter.
The results of the study of the longitudinal hadronic shower development are
presented.
The data have been taken in the H8 beam line of the CERN SPS using pions 
of 10 -- 300 GeV.
\end{abstract}

\newpage
\section{Introduction}
The key question of calorimetry generally and hadronic calorimetry in 
particular is the energy reconstruction.
This question is especially important when a hadronic calorimeter have a 
complex structure being a combined calorimeter.
Such is the combined calorimeter with the electromagnetic and hadronic 
compartments of the ATLAS detector 
\cite{atcol94,TILECAL96,LARG96}.
In this paper we describe the non-parametrical method of the energy 
reconstruction for a combined calorimeter, which called the $e/h$ method, 
and demonstrate its performance on the basis of the test beam data of the 
ATLAS combined prototype calorimeter.
For the energy reconstruction and description of the longitudinal 
development of a hadronic shower it is necessary to know the $e/h$ ratios, 
the degree of non-compensation, of these  calorimeters. 
As to the ATLAS Tile barrel calorimeter there is the detailed information 
about the $e/h$ ratio presented in
\cite{TILECAL96,ariztizabal94,juste95,budagov96-72,kulchitsky99-12}.
But as to the liquid argon electromagnetic calorimeter such information
is practically absent.
The aim of the present work is also to develop the method and to determine
the value of the $e/h$ ratio of the electromagnetic compartment.

One of the important questions of hadron calorimetry is the question
of the longitudinal development of hadronic showers.
This question is especially important for a combined calorimeter.
This work is also devoted to the study of the longitudinal hadronic shower 
development in the ATLAS barrel combined prototype calorimeter.
    
This work has been performed on the basis of the 1996 combined test beam 
data \cite{comb96,cobal98}.
The data have been taken in the H8 beam line of the CERN SPS using pions 
of  10, 20, 40, 50, 80, 100, 150 and 300 GeV.

\section{Combined Calorimeter}
\begin{figure}[tbph]
\begin{center}
\mbox{\epsfig{figure=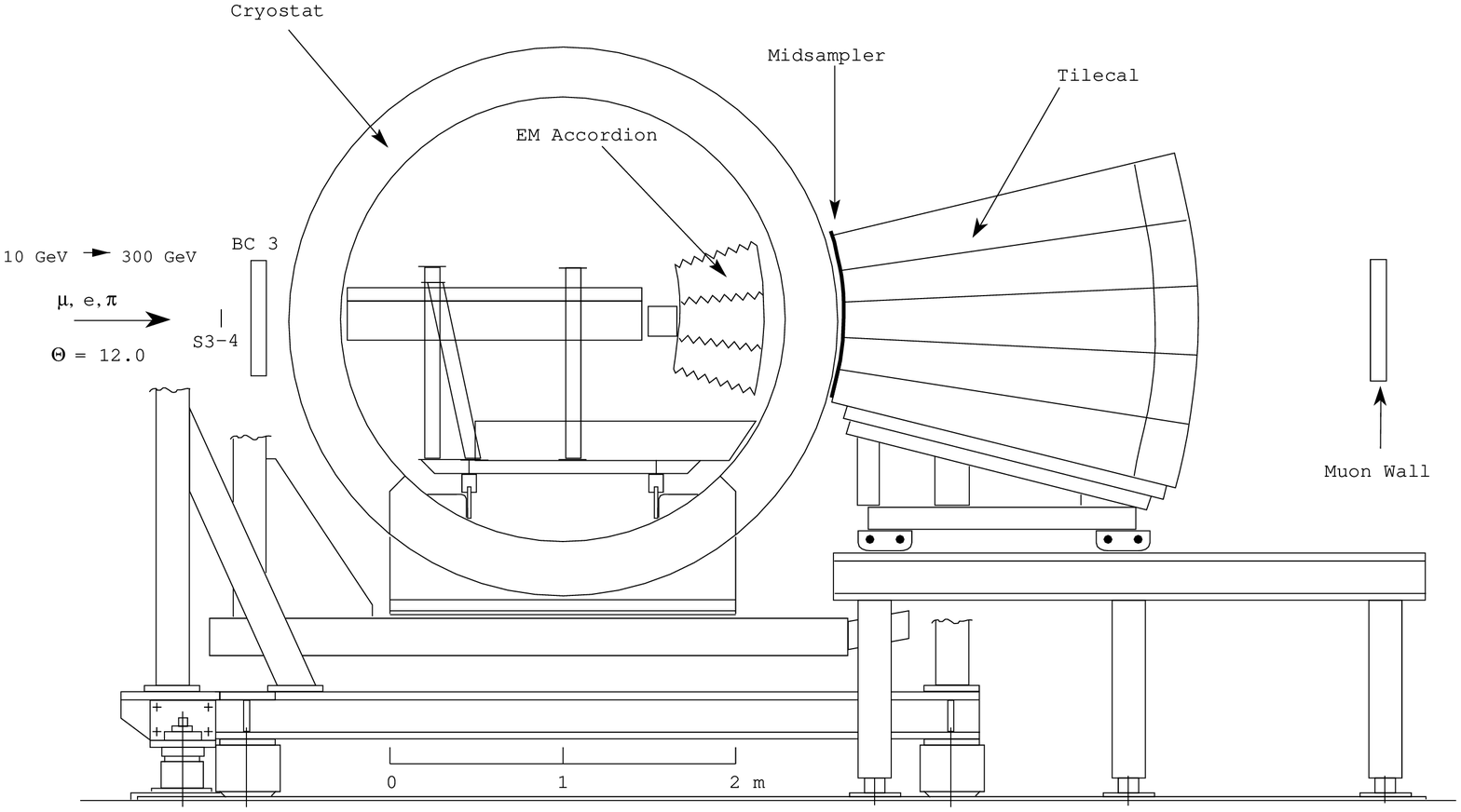,width=0.9\textwidth,height=0.35\textheight}} 
\end{center}
 \caption{
        Test beam setup for the combined LAr and Tile calorimeters run.}
\label{fv1}
\end{figure}

The combined calorimeter prototype setup has been made consisting of the 
LAr electromagnetic calorimeter prototype inside the cryostat and 
downstream the Tile calorimeter prototype as shown in Fig.\ \ref{fv1}. 
The two calorimeters have been placed with their central axes at an angle 
to the beam of $12^\circ$.
At this angle the two calorimeters have an active thickness of 10.3 
$\lambda_I$.
Beam quality and geometry were monitored with a set of beam wire chambers 
BC1, BC2, BC3 and trigger hodoscopes placed upstream of the LAr cryostat.
To detect punchthrough particles and to measure the effect of longitudinal 
leakage a ``muon wall'' consisting of 10 scintillator counters (each 2 cm 
thick) was located behind the calorimeters at a distance of about 1 metre.

\subsection{Electromagnetic Calorimeter}
The electromagnetic LAr calorimeter  prototype consists of a stack of three 
azimuthal modules, each one spanning $9^\circ$ in azimuth and extending over
2 m along the Z direction.
The calorimeter structure is defined by 2.2 mm thick steel-plated lead 
absorbers folded to an accordion shape and separated by 3.8 mm gaps filled 
with liquid argon.
The signals are collected by Kapton electrodes located in the gaps.
The calorimeter extends from an inner radius of 131.5 cm to an outer radius
of 182.6 cm, representing (at $\eta = 0$) a total of 25 radiation lengths 
($X_0$), or 1.22 interaction lengths ($\lambda_I$) for protons.
The calorimeter is longitudinally segmented into three compartments of 
$9\ X_0$, $9\ X_0$ and $7\ X_0$, respectively.
More details about this prototype can be found in 
\cite{atcol94,ccARGON}.

The cryostat has a cylindrical form with 2 m internal diameter, filled with 
liquid argon, and is made out of a 8 mm thick inner stainless-steel vessel,
isolated by 30 cm of low-density foam (Rohacell), itself protected by a 
1.2 mm thick aluminum outer wall.

\subsection{Hadronic Calorimeter}
The  hadronic Tile calorimeter is a sampling device using steel as the 
absorber and scintillating tiles as the active material 
\cite{TILECAL96}.
The innovative feature of the design is the orientation of the tiles which 
are placed in planes perpendicular to the Z direction 
\cite{gild91}.
For a better sampling homogeneity the 3 mm thick scintillators are staggered
in the radial direction.
The tiles are separated along  Z  by 14 mm of steel, giving a 
steel/scintillator volume ratio of 4.7.
Wavelength shifting fibres (WLS) running radially collect light from the 
tiles at both of their open edges.
The hadron calorimeter prototype consists of an azimuthal stack of five 
modules.
Each module covers $2\pi/64$\ in azimuth and extends 1 m along the Z 
direction, such that the front face covers $100\times20$\ cm$^2$. 
The radial depth, from an inner radius of 200 cm to an outer radius of 
380 cm, accounts for 8.9 $\lambda$\ at $\eta = 0$\  (80.5 $X_0$).
Read-out cells are defined by grouping together a bundle of fibres into one 
photomultiplier (PMT).
Each of the 100 cells is read out by two PMTs and is fully projective in 
azimuth (with $\Delta \phi = 2\pi/64 \approx 0.1$), while the segmentation 
along the Z axis is made by grouping fibres into read-out cells spanning
$\Delta Z = 20$\ cm ($\Delta \eta \approx 0.1$) and is therefore not 
projective.
Each module is read out in four longitudinal segments (corresponding to 
about 1.5, 2, 2.5 and 3 $\lambda_I$\ at $\eta = 0$).
More details of this prototype can be found in
\cite{atcol94,ccNIM,ccrd34rep94,shower98,budagov-97-127}.

\subsection{Data Selection}
The data have been taken in the H8 beam line of the CERN SPS using pions of 
10, 20, 40, 50, 80, 100, 150 and 300 GeV.
We applied some similar to 
\cite{comb96,cobal98} 
cuts to eliminate the non-single track pion events, the beam halo, the 
events with an interaction before the LAr calorimeter, the electron and 
muon events.
The set of cuts is the following:
the single-track pion events were selected by requiring the pulse height of 
the beam scintillation counters and the energy released in the presampler of
the electromagnetic calorimeter to be compatible with that for a single 
particle; 
the beam halo events were removed with appropriate cuts on the horizontal
and vertical positions of the incoming track impact point and the space 
angle with respect to the beam axis as measured with the beam chambers; 
a cut on the total energy rejects incoming muon. 

\section{$e/h$ Method of Energy Reconstruction}
The response, $R$, of a calorimeter to a hadronic shower is the sum of the 
contributions from the electromagnetic, $E_e$, and hadronic, $E_h$, parts of
the incident energy $E$
\cite{wigmans88,groom89}:
\begin{equation}
        R = 
                  e \cdot E_e + h \cdot E_h 
            =     e \cdot E \cdot 
                 (f_{\pi^0} + (h / e) \cdot ( 1 - f_{\pi^0})) \ , 
\label{ev9}
\end{equation}
\begin{equation}
        E = E_e + E_h \ ,
\label{ev18}
\end{equation}
where $e$ ($h$) is the energy independent coefficient of transformation of 
the electromagnetic (pure hadronic, low-energy hadronic activity) energy to 
response, $f_{\pi^0} = E_e / E$ is the fraction of electromagnetic 
energy.
From this
\begin{equation}
        E = \Bigl( \frac{e}{\pi} \Bigr) \cdot \frac{R}{e} \ ,
\label{ev16}
\end{equation}
and  
\begin{equation}
        \frac{e}{\pi} = 
                \frac{e/h}{1+(e/h-1)f_{\pi^0}} \ .
\label{ev10}
\end{equation}

For a combined calorimeter the incident energy deposits into the LAr 
compartment, $E_{LAr}$, the Tile calorimeter compartment, $E_{Tile}$, and 
into the passive material between the LAr and Tile calorimeters, $E_{dm}$,
\begin{equation}
        E =  E_{LAr} + E_{dm} + E_{Tile} \ . 
\label{ev13}
\end{equation}

Using the expressions (\ref{ev16}) -- (\ref{ev13}) the following equation 
for the energy reconstruction has been derived:
\begin{equation}
        E =    \sum_i c_i \cdot\Bigl(\frac{e}{\pi}\Bigr)_i \cdot R_i   
               =c_{LAr}\cdot\Bigl(\frac{e}{\pi}\Bigr)_{LAr}\cdot R_{LAr} 
               +E_{dm}
            +c_{Tile}\cdot\Bigl(\frac{e}{\pi}\Bigr)_{Tile}\cdot R_{Tile}\ , 
\label{ev7}
\end{equation}
where $i = LAr,\ dm,\ Tile$; $c_i =1/e_i$; $(e/\pi)_i$ are from equation 
(\ref{ev10}) and
\begin{equation}
        f_{\pi^0,\ LAr} = k \cdot \ln{(E)} \ ,  
\label{efpl}
\end{equation} 
\begin{equation}
        f_{\pi^0,\ Tile} = k \cdot \ln{(E_{Tile})} \ ,
\label{fpt}
\end{equation}
where $E_{Tile}=c_{Tile}\cdot (e/\pi)_{Tile}\cdot R_{Tile}$ and $k = 0.11$.
Note, that for $\approx 70\%$ of events an energy in Tile calorimeter is 
approximately equal to beam energy because a hadronic shower began in the 
hadron calorimeter. 

The term, which accounts for the energy loss in the dead material between 
the LAr and Tile calorimeters, $E_{dm}$, is taken to be proportional to the 
geometrical mean of the energy released in the third depth of the 
electromagnetic compartment
($E_{LAr,\  3} = c_{LAr}  \cdot (e/\pi)_{LAr}  \cdot R_{LAr,\ 3}$) 
and the first depth of the hadronic compartment 
($E_{Tile,\ 1} = c_{Tile} \cdot (e/\pi)_{Tile} \cdot R_{Tile,\ 1}$)
\begin{equation}
\label{ev19}
        E_{dm} = c_{dm} \cdot \sqrt{E_{LAr, 3} \cdot E_{Tile, 1}}
\end{equation}
similar to
\cite{comb96,combined94}.
The validity of this approximation has been tested by the Monte Carlo 
simulation and by the study of the correlation between the energy released
in the midsampler and the cryostat energy deposition 
\cite{cobal98,bosman99,atcol99}. 
The value of $c_{dm} = 0.31$ obtained on the basis of the results of the 
Monte Carlo simulation is used.

In order to use the equation (\ref{ev7}) it is necessary to know the values
of the following constants: $c_{LAr}$, $c_{dm}$, $c_{Tile}$, $(e/h)_{LAr}$, 
$(e/h)_{Tile}$, some of which are $c_{LAr}=1/e_{LAr}=1.1$ 
\cite{comb96,cobal98}, 
$(e/h)_{Tile}=1.3\pm0.03$ \cite{budagov96-72}.
The determination of the other constants 
($c_{dm}$, $c_{Tile}$, $(e/h)_{LAr}$) is given below.

\subsection{$c_{Tile}$ Constant}
For the determining of the $c_{Tile}$ constant the following procedure was 
applied.   
We selected the events which start to shower only in the hadronic 
calorimeter.
To select these events the energies deposited in each sampling of the LAr 
calorimeter and in the midsampler are required to be compatible with that 
of a single minimum ionization particle. 

The following expression for the normalized hadronic response have been 
used  \cite{groom89}:
\begin{equation}
\frac{R_{Tile}^{c}}{E_{beam}} =
\frac{1+((e/h)_{Tile}-1)\cdot f_{\pi^0,\ Tile}}{c_{Tile}
\cdot (e/h)_{Tile}}\ , 
\label{ev14}
\end{equation}
where
\begin{equation}
        R_{Tile}^{c} = 
                R_{Tile} + \frac{c_{LAr}}{c_{Tile}} \cdot R_{LAr}
\label{ev17}
\end{equation}
is the Tile calorimeter response corrected on the energy loss in the
LAr calori\-meter, $f_{\pi^0, Tile}$ is determined by the formula 
(\ref{fpt}).
The value of $c_{Tile}$ obtained by fitting is equal to $0.145\pm0.002$.

\subsection{$e/h$ ratio of Electromagnetic Compartment}
Using the expression (\ref{ev7}) the value of the $(e / \pi)_{LAr}$ ratio 
can be obtained
\begin{equation}
        \Bigl( \frac{e}{\pi} \Bigr)_{LAr} =
        \frac{E_{beam} - E_{dm} - E_{Tile}}{c_{LAr} \cdot R_{LAr}} \ . 
\label{ev1}
\end{equation}
The $(e/h)_{LAr}$ ratio and the function $f_{\pi^0, LAr}$ (\ref{efpl}) can 
be inferred from the energy dependent $(e/\pi)_{LAr}$ ratios.

For this case we select the events with the well developed hadronic showers
in the electromagnetic calorimeter. Than mean that energy depositions were 
required to be more than 10\% of the beam energy in the electromagnetic 
calorimeter and less than 70\% in the hadronic calorimeter. 

\begin{figure*}[tbph]
\vspace*{-20 mm}
\hspace*{-20 mm}
\begin{center}   
\begin{tabular}{cccc}
\mbox{\epsfig{figure=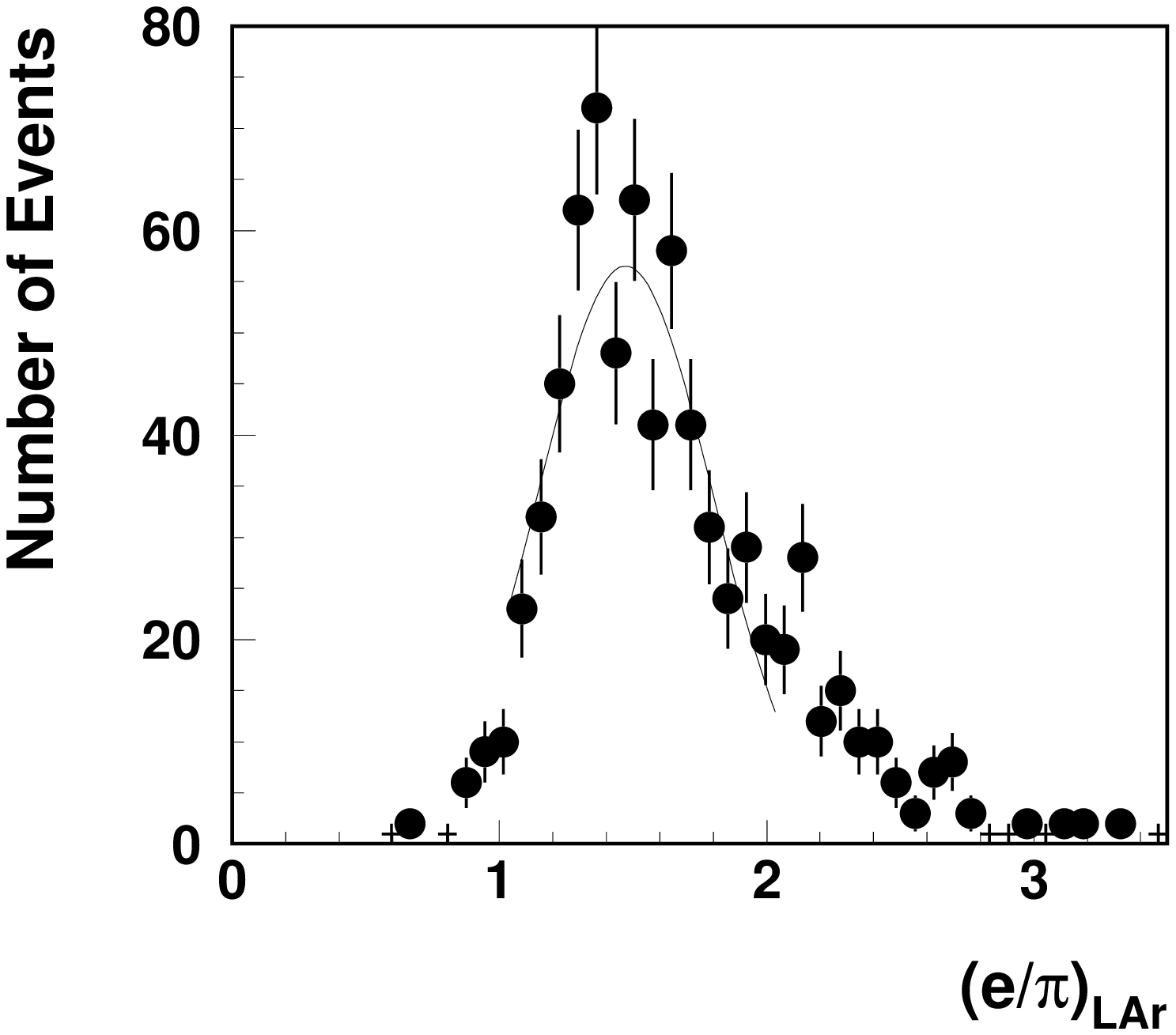,width=0.2\textwidth,height=0.5\textheight}}
& 
\mbox{\epsfig{figure=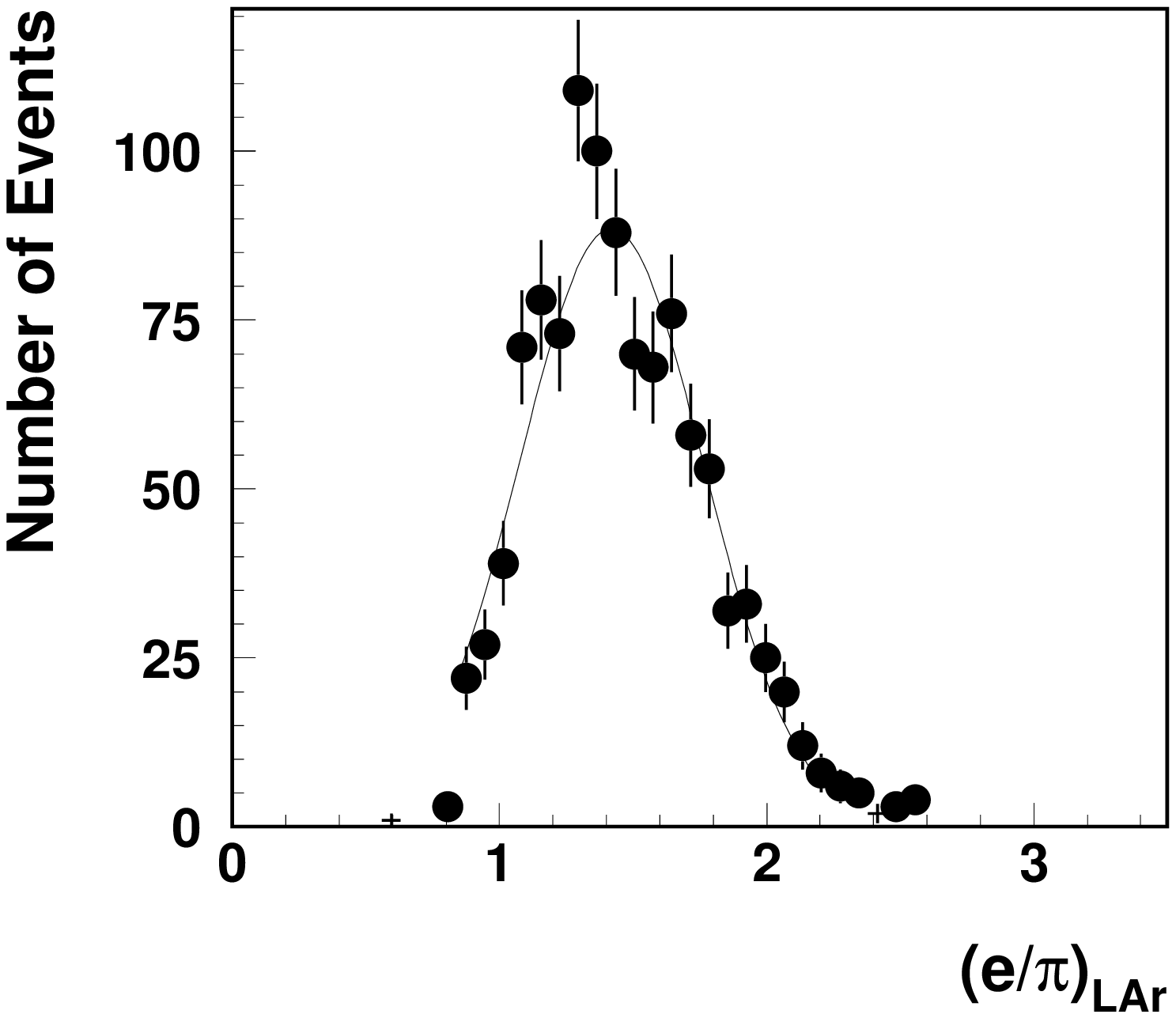,width=0.21\textwidth,height=0.5\textheight}}
&
\mbox{\epsfig{figure=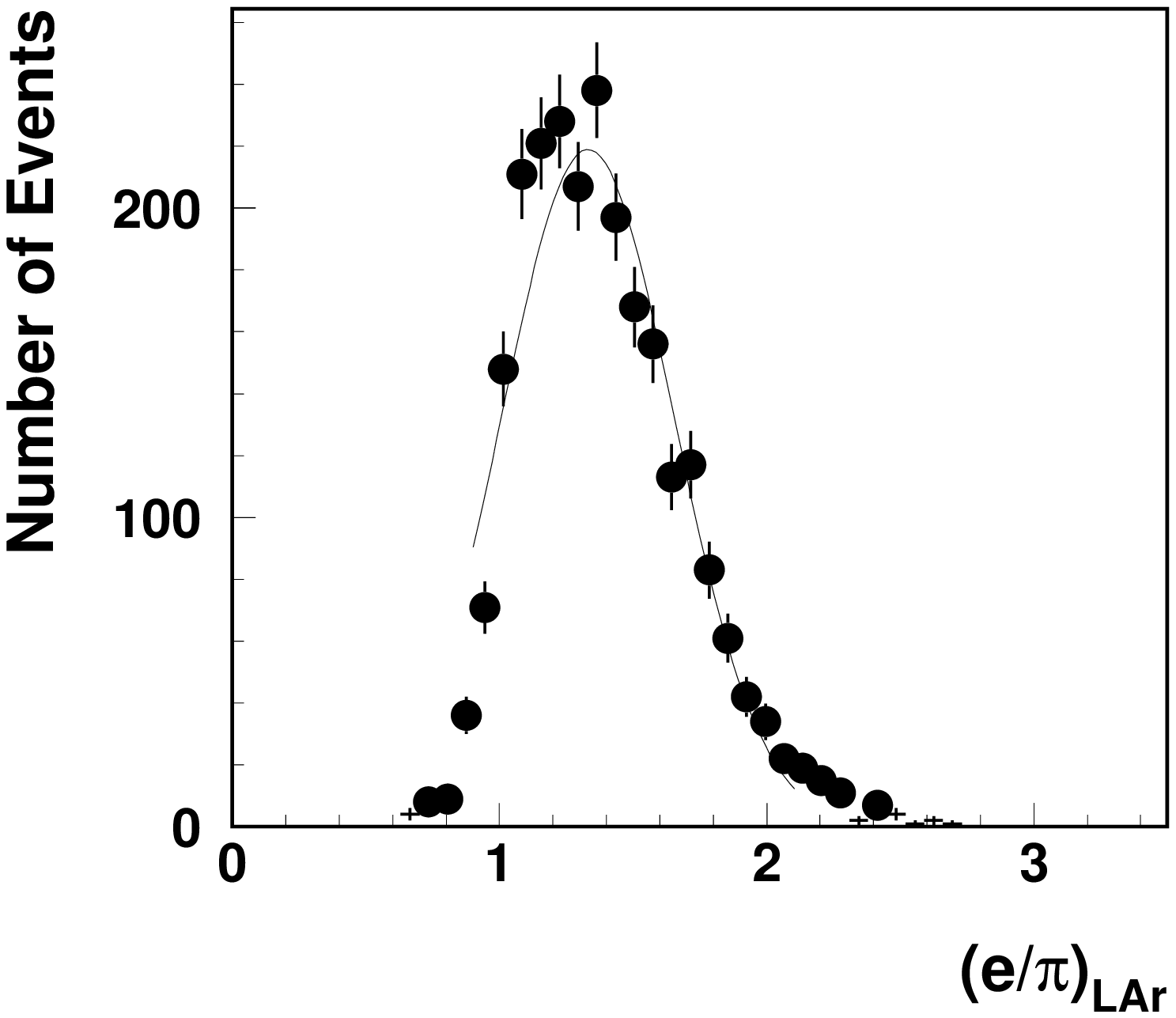,width=0.21\textwidth,height=0.5\textheight}}
& 
\mbox{\epsfig{figure=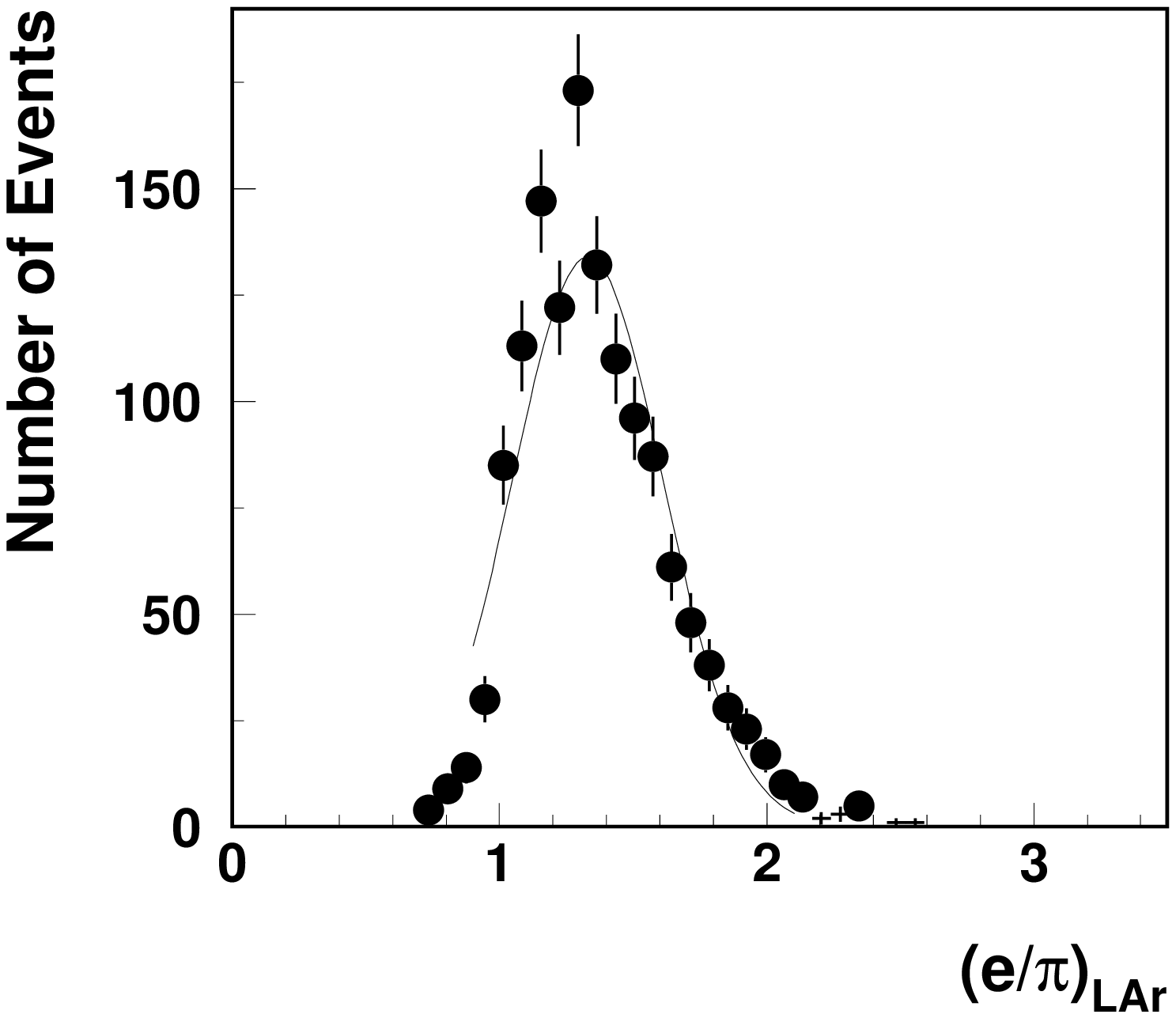,width=0.21\textwidth,height=0.5\textheight}}
\\[-15mm]
\mbox{\epsfig{figure=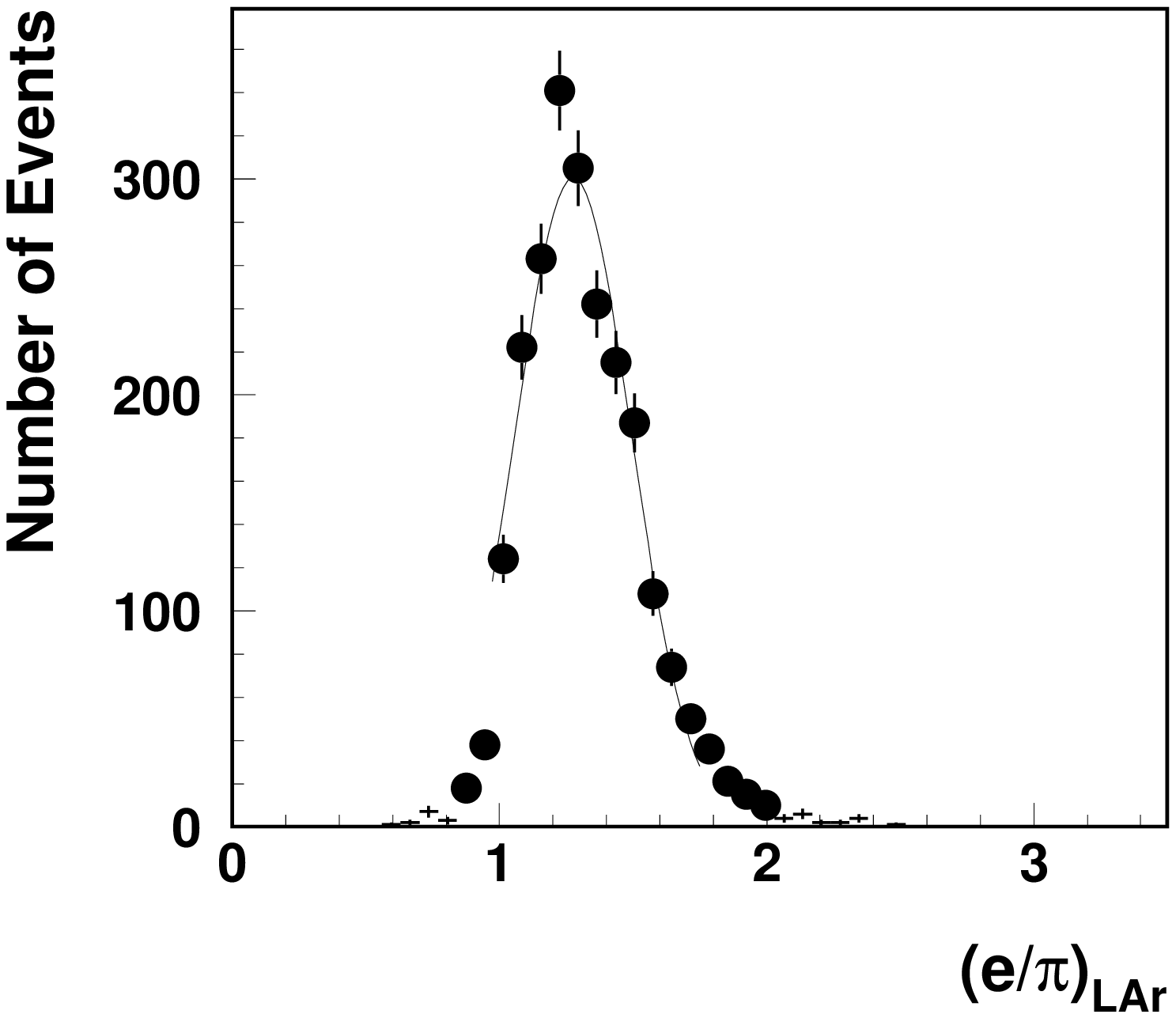,width=0.21\textwidth,height=0.5\textheight}}
& 
\mbox{\epsfig{figure=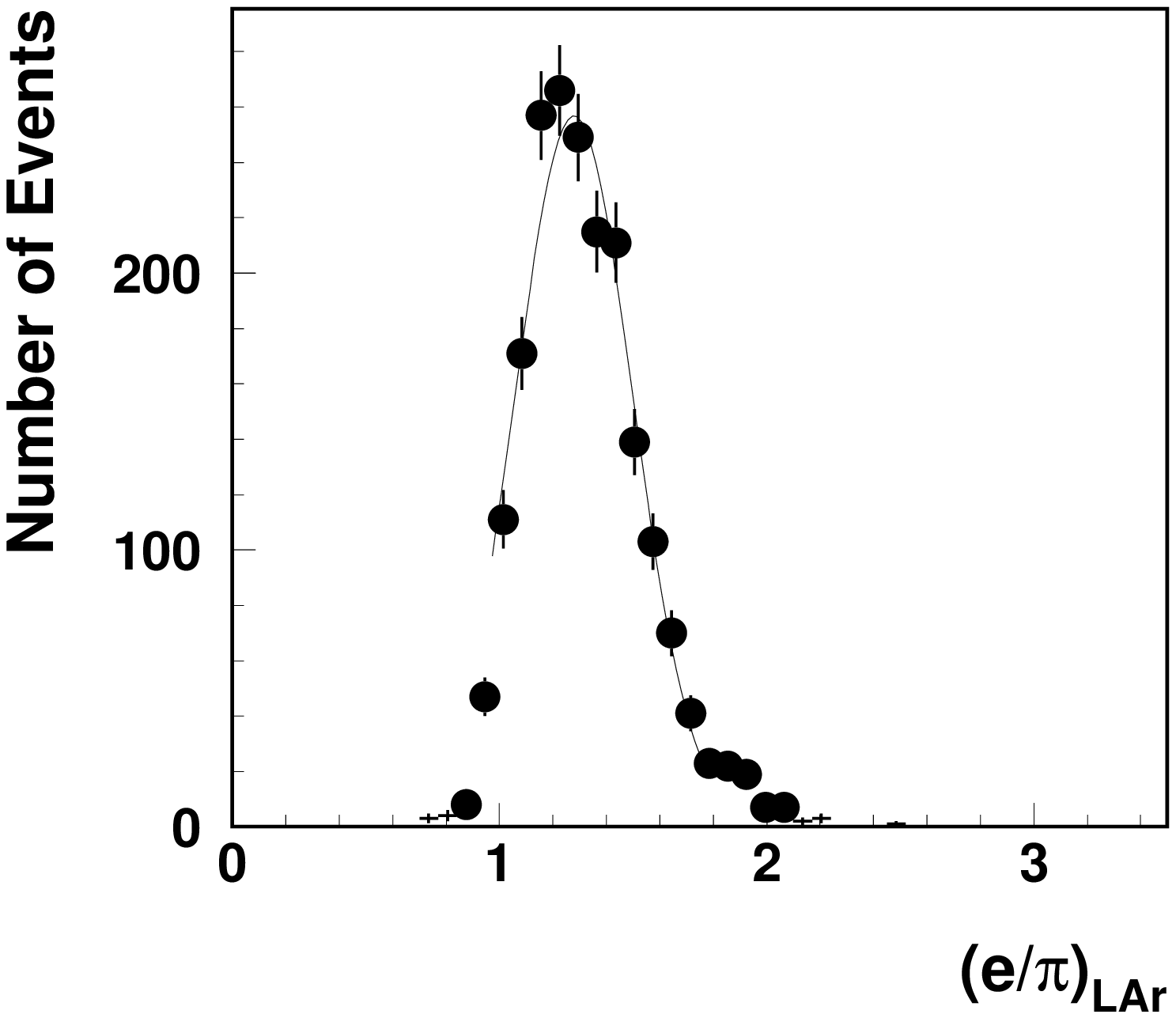,width=0.21\textwidth,height=0.5\textheight}}
&
\mbox{\epsfig{figure=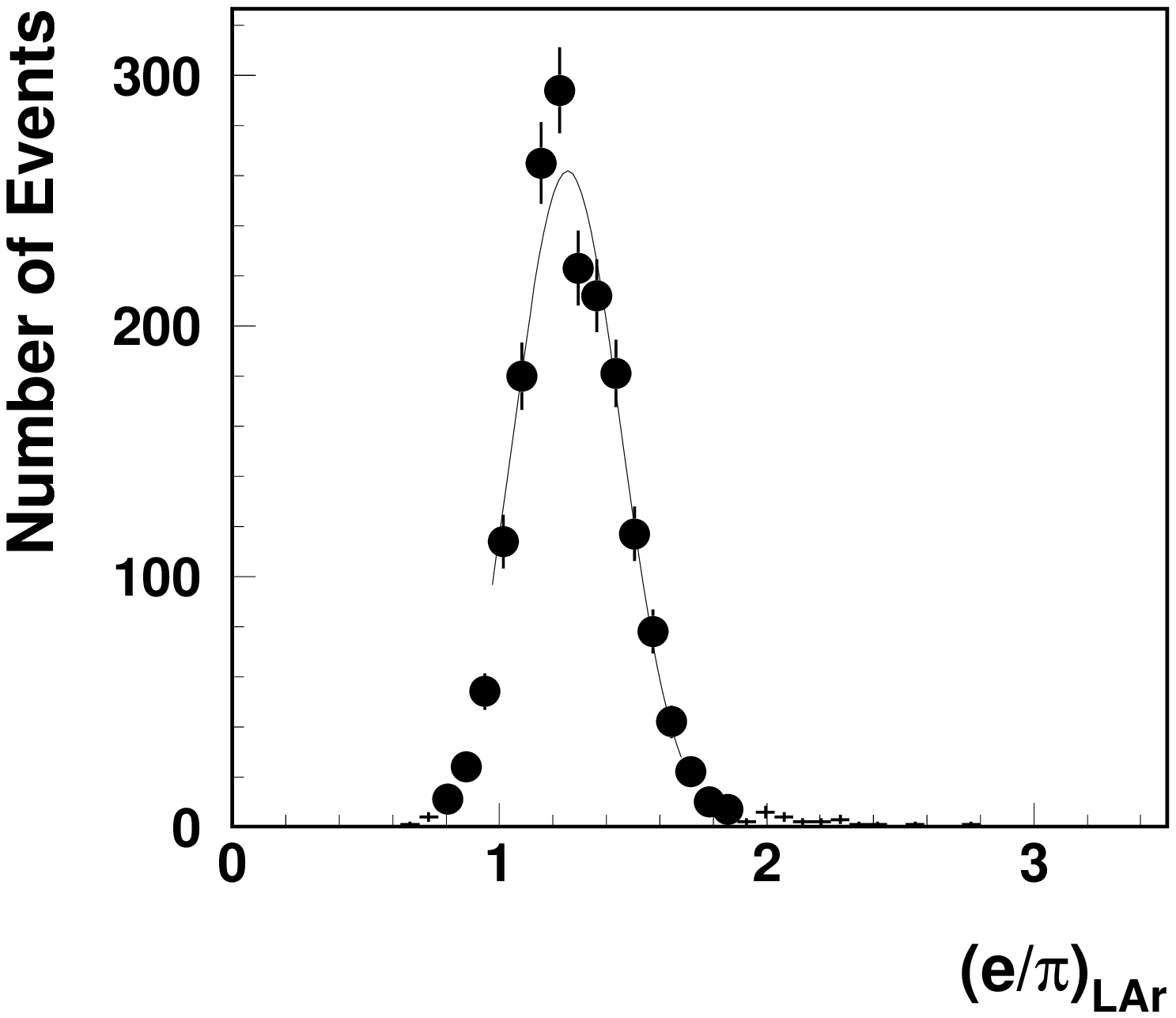,width=0.22\textwidth,height=0.5\textheight}}
&
\mbox{\epsfig{figure=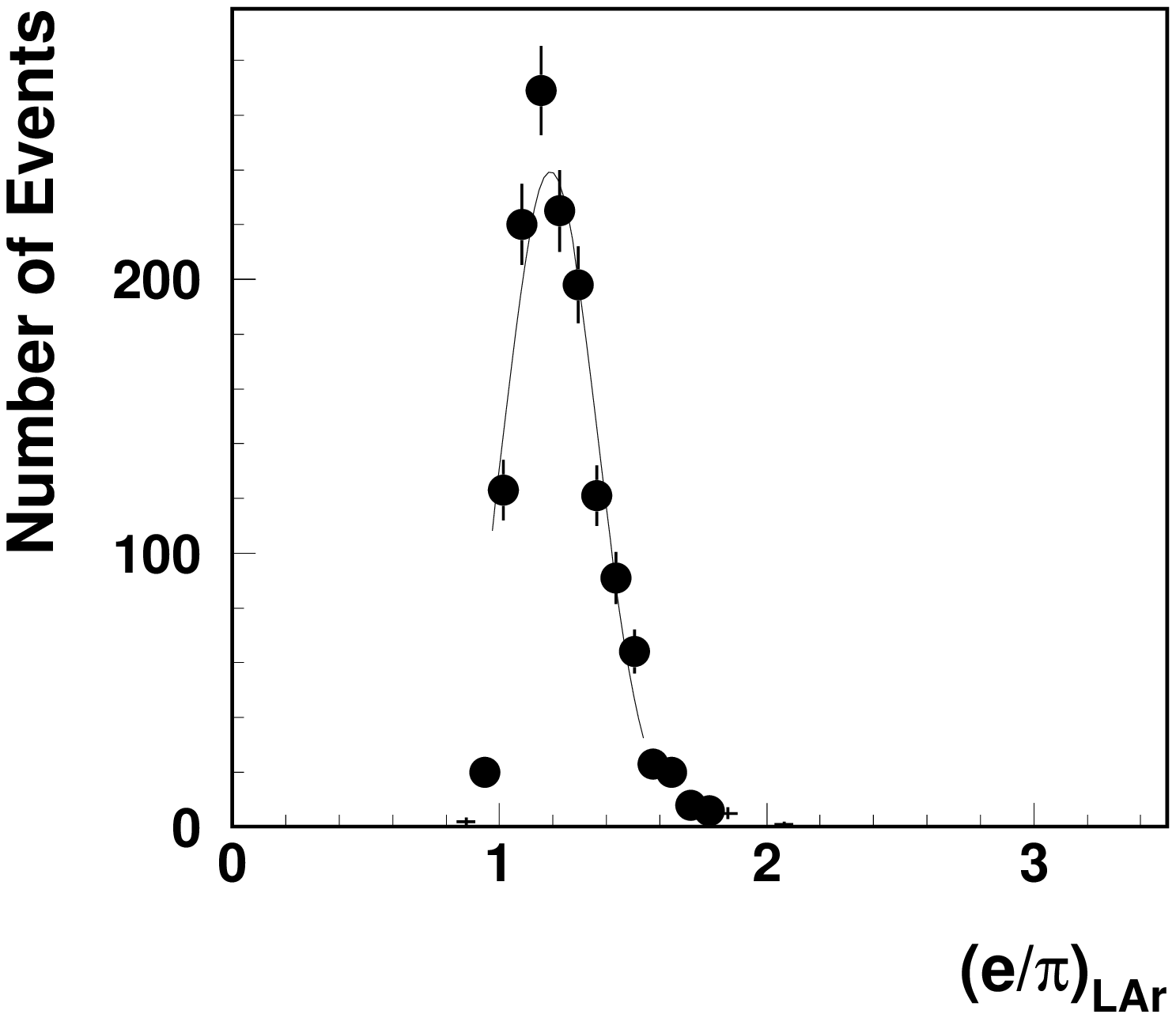,width=0.22\textwidth,height=0.5\textheight}}
\\[-10mm]
\end{tabular}
\end{center}
       \caption{The distributions of the $(e / \pi)_{LAr}$ ratio 
                for $E_{beam}$ = 10, 20, 40, 50 GeV
                (top row, left to right) 
                and $E_{beam}$ = 80, 100, 150, 300 GeV
                (bottom row, left to right). 
       \label{fv2}}
\end{figure*}

\begin{figure*}[tbph]
\begin{center}   
\begin{tabular}{c}
\mbox{\epsfig{figure=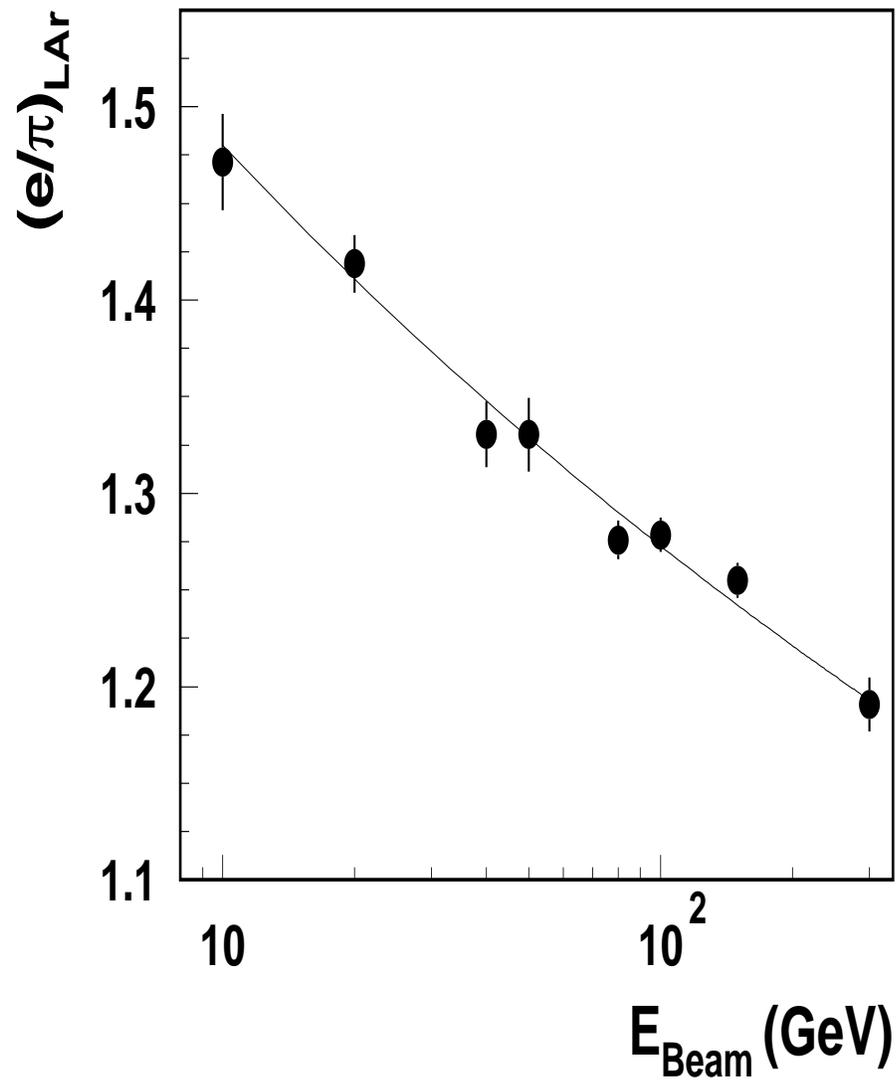,width=0.95\textwidth,height=0.9\textheight}}
\\[-10mm]
\end{tabular}
\end{center}
       \caption{
        The mean value of the $(e/\pi)_{LAr}$ ratio as a function 
        of the beam energy.
        The curve  is the result of a fit of equation (\ref{ev10}).
      \label{fv3}}
\end{figure*}

Fig.\ \ref{fv2} shows the distributions of the $(e/\pi)_{LAr}$ ratio 
derived by formula (\ref{ev1}) for different energies.
The mean values of these distributions are shown in Fig.\ \ref{fv3} as a 
function of the beam energy.
The fit of this distribution by the expression (\ref{ev10}) for LAr 
calorimeter yields $(e/h)_{LAr}=1.74\pm0.04$ and $k = 0.108\pm0.004$. 
The quoted errors are the statistical ones obtained from the fit.
The systematic error on the $(e/h)_{LAr}$ ratio, which is a consequence of 
the uncertainties in the input constants used in the equation (\ref{ev1}), 
is estimated to be $\pm0.04$.

Wigmans  showed 
\cite{wigmans91} 
that the $e/h$ ratio for non-uranium calorimeters with high-Z absorber 
material is satisfactorily described by the formula:
\begin{equation}
        \frac{e}{h}  = \frac{e/mip}{0.41 + 0.12 \cdot n/mip}
\label{wig}
\end{equation}
in which $e/mip$ and $n/mip$ 
represent the calorimeter response to e.m.\
showers and to MeV-type neutrons, respectively.
These responses are normalized to the one for minimum ionizing particles.
The Monte Carlo calculated $e/mip$ and $n/mip$ values for the R{\&}D3 Pb-LAr
electromagnetic calorimeter are $e/mip = 0.78$ \cite{wigmans91} 
and $n/mip < 0.5$ \cite{wigmans91} leading to 
$(e/h)_{LAr} > 1.66$.
Our measured value of the $(e/h)_{LAr}$ ratio agrees with this prediction.  

\subsection{Iteration Procedure}
For the energy reconstruction by the formula (\ref{ev7}) it is necessary to 
know the $(e/\pi)_{Tile}$ ratio and the reconstructed energy itself.
Therefore, the iteration procedure has been developed.
Two iteration cycles were made: the first one is devoted to the 
determination of the $(e/\pi)_{Tile}$ ratio and the second one is the 
energy reconstruction itself.

The expression (\ref{ev10}) for the $(e/\pi)_{Tile}$ ratio can be written as
\begin{equation}
        \Bigl(\frac{e}{\pi}\Bigr)_{Tile} = 
                \frac{(e/h)_{Tile}}{1+((e/h)_{Tile}-1) 
                \cdot k 
              \cdot \ln{(c_{Tile} \cdot (e/\pi)_{Tile} \cdot R_{Tile})}}\ .
\label{epti}
\end{equation} 
As the first approximation, the value of $(e/\pi)_{Tile}$ is calculated 
using the equation (\ref{epti})
where in the right side of this equation we used $(e/\pi)_{Tile} = 1.13$ 
corresponding to $f_{\pi^0, Tile} = 0.5 = 0.11 \ln{(100\ GeV)}$.
The iteration process is stopped when the convergence criterion
$\mid (e/\pi)_{Tile}^{\nu + 1} - (e / \pi)_{Tile}^{\nu }\mid 
     /(e/\pi)_{Tile}^{\nu} < \epsilon$,    
where ($\nu = 0,\ 1, \ldots$), is satisfied. 

As the first approximation in the iteration cycle for the energy 
reconstruction, the value of $E$ is calculated using the equation 
(\ref{ev7}) with the $(e/\pi)_{Tile}$ ratio obtained in the first 
iteration cycle and $(e/\pi)_{LAr}$ from equation (\ref{ev10}) where 
in the right side of this equation we used $(e/\pi)_{LAr} = 1.27$ 
corresponding to $f_{\pi^0, LAr} = 0.5 = 0.11 \ln{(100\ GeV)}$.
The convergence criterion is
$\mid E^{\nu +1} - E^{\nu} \mid / E^{\nu} < \epsilon$.

For both these case the iteration values are under the logarithmic 
function that mean that iteration procedure will be very fast. 

The average numbers of iterations $<N_{it}>$ for the various beam energies 
needed to receive the given value of accuracy $\epsilon$ have been 
investigated.
It turned out, it is sufficiently only the first approximation for 
achievement, on average, of convergence with an accuracy of $\epsilon =1\%$
for energies 80 -- 150 $GeV$ and it is necessary to perform only one 
iteration for the energies at 10 -- 50 $GeV$ and 300 $GeV$.

\begin{figure*}[tbph]
\begin{center}   
\begin{tabular}{c}
\epsfig{figure=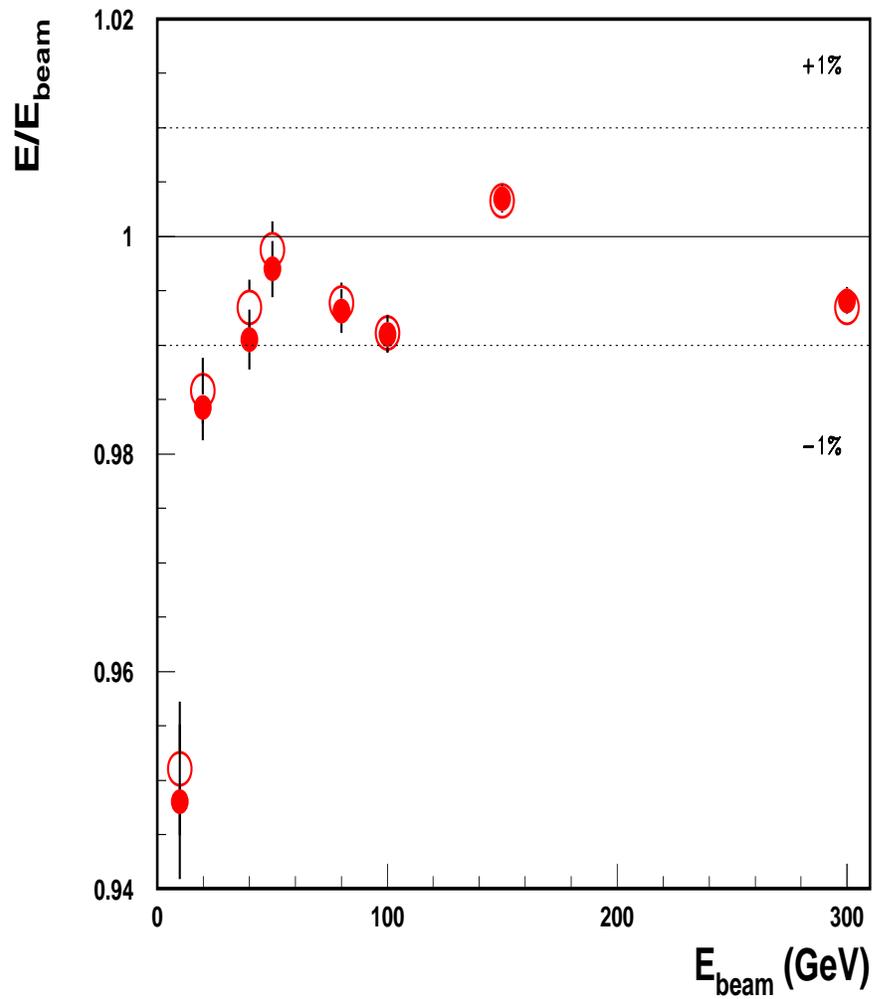,width=0.95\textwidth,height=0.9\textheight} 
\\[-10mm]
\end{tabular}
\end{center}
       \caption{ 
         Energy linearity as a function of the beam energy for 
         the $e/h$ method  obtained using the iteration procedure 
         with $\epsilon = 0.1\%$ (black circles) and the first 
         approximation (open circles).
       \label{f03-0}}
\end{figure*}

\begin{figure*}[tbph]
\begin{center}   
\begin{tabular}{c}
\epsfig{figure=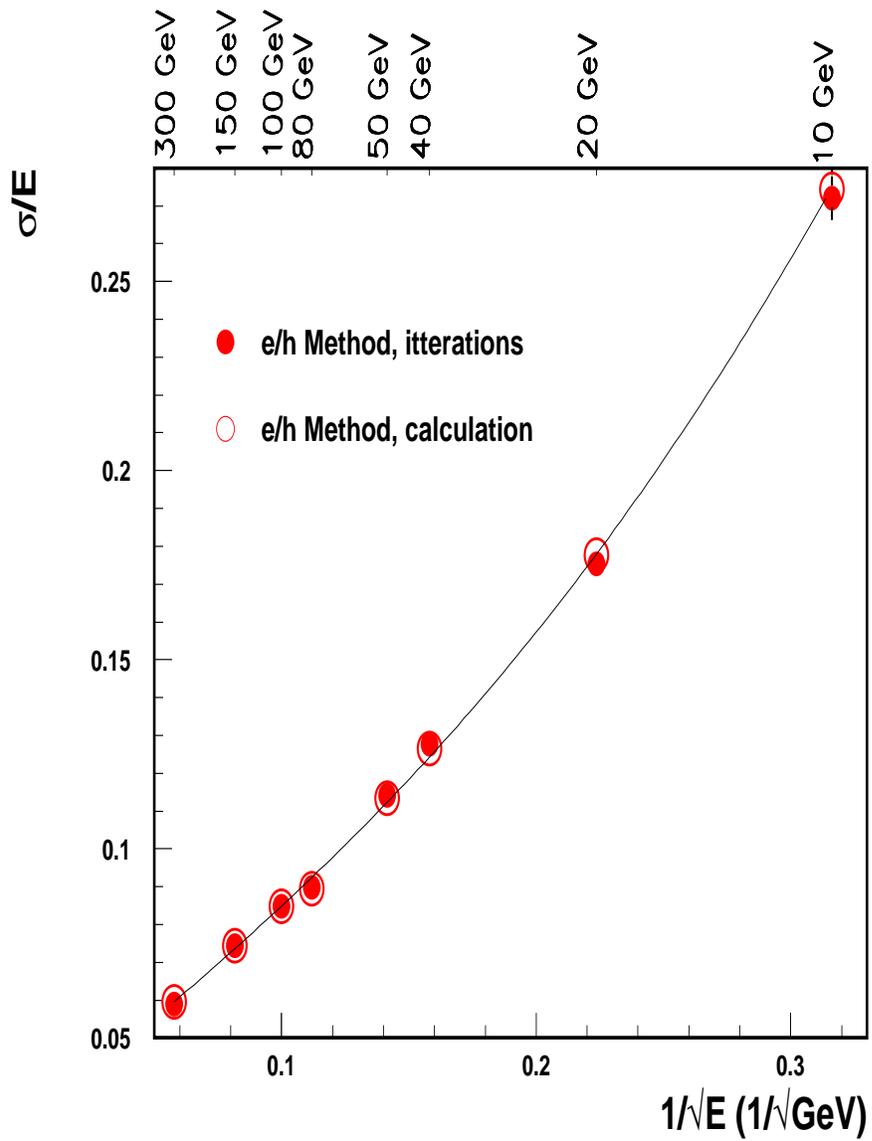,width=0.95\textwidth,height=0.9\textheight} 
\\[-10mm]
\end{tabular}
\end{center}
       \caption{  
         The fractional energy resolutions obtained 
         with the $e/h$ method (black circles), 
         and the first approximation (open circles).          
       \label{f05-0}}
\end{figure*}

We specially investigated the accuracy of the first approximation of energy.
Fig.\ \ref{f03-0} shows the comparison between the energy linearities, the 
mean values of $E / E_{beam}$, obtained using the iteration procedure with 
$\epsilon = 0.1\%$ (black circles) and the first approximation of energy 
(open circles).
Fig.\ \ref{f05-0} shows the comparison between the energy resolutions 
obtained using these two approaches.
As can be seen, the compared values are consistent within errors.

The suggested algorithm of the energy reconstruction can be used for
the fast energy reconstruction in the first level trigger.

\subsection{Energy Spectra}
\begin{figure*}[tbph]
\vspace*{-20 mm}
\hspace*{-20 mm}
\begin{center}   
\begin{tabular}{cccc}
\epsfig{figure=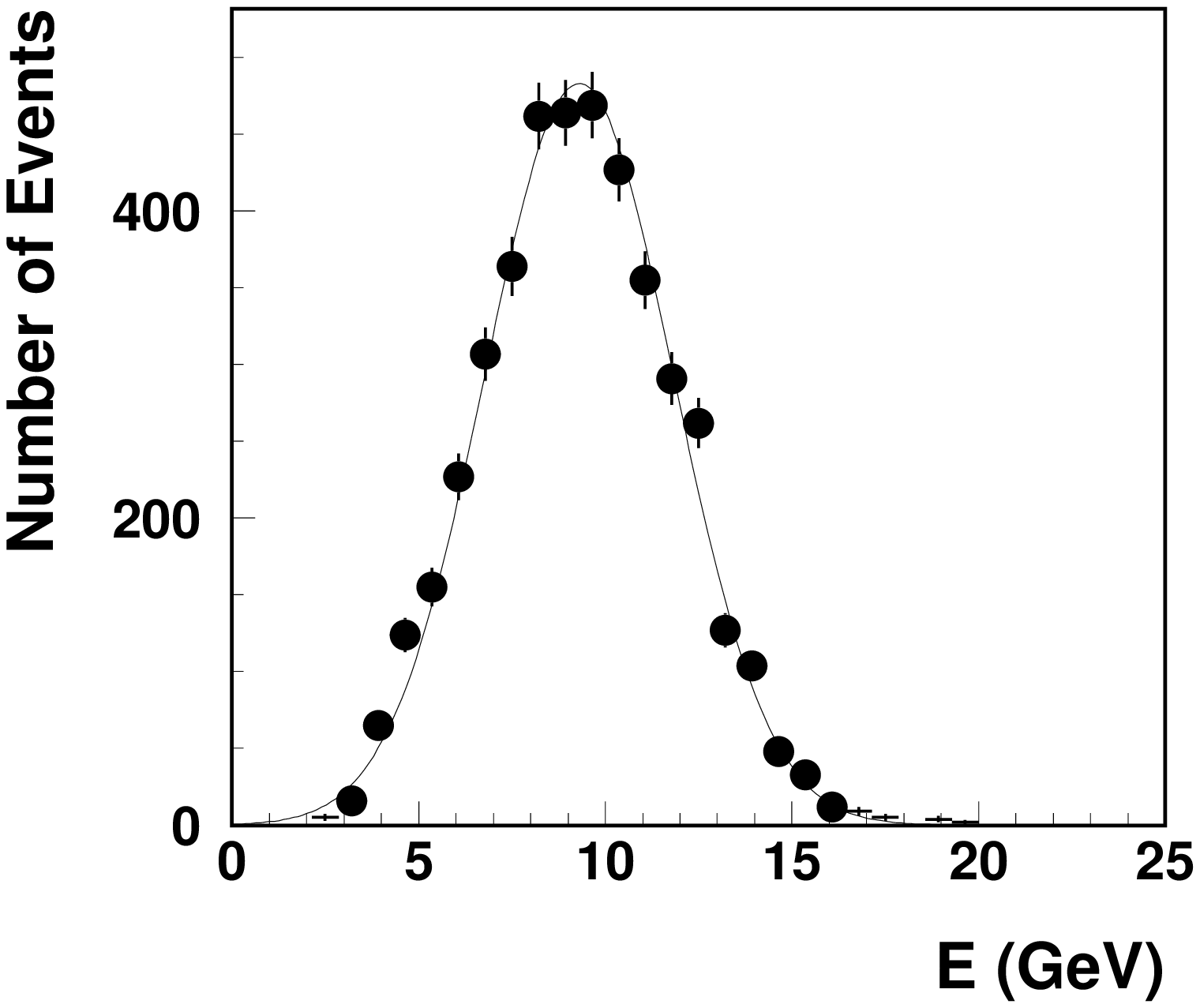,width=0.21\textwidth,height=0.5\textheight}
&
\epsfig{figure=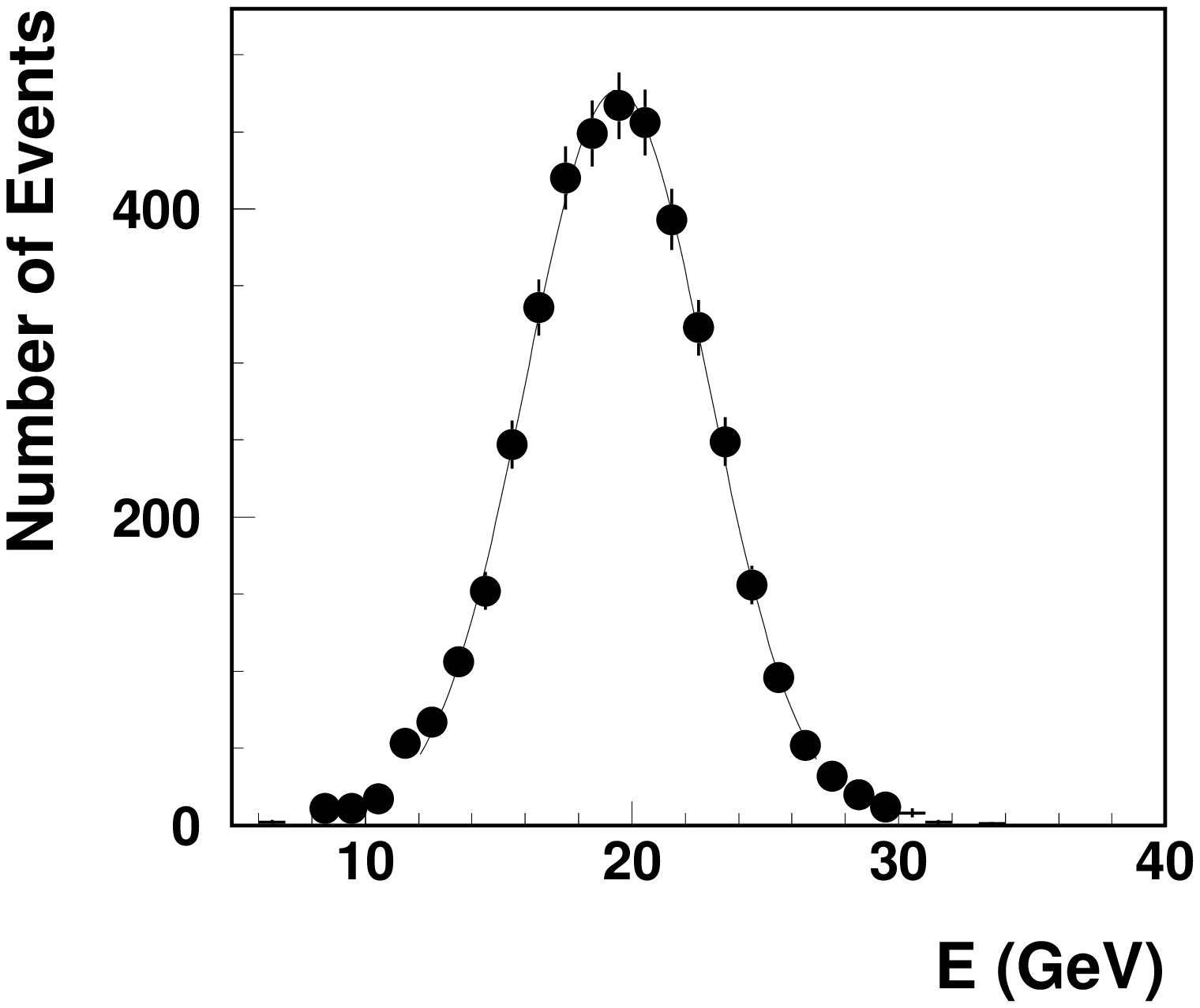,width=0.21\textwidth,height=0.5\textheight} 
&
\epsfig{figure=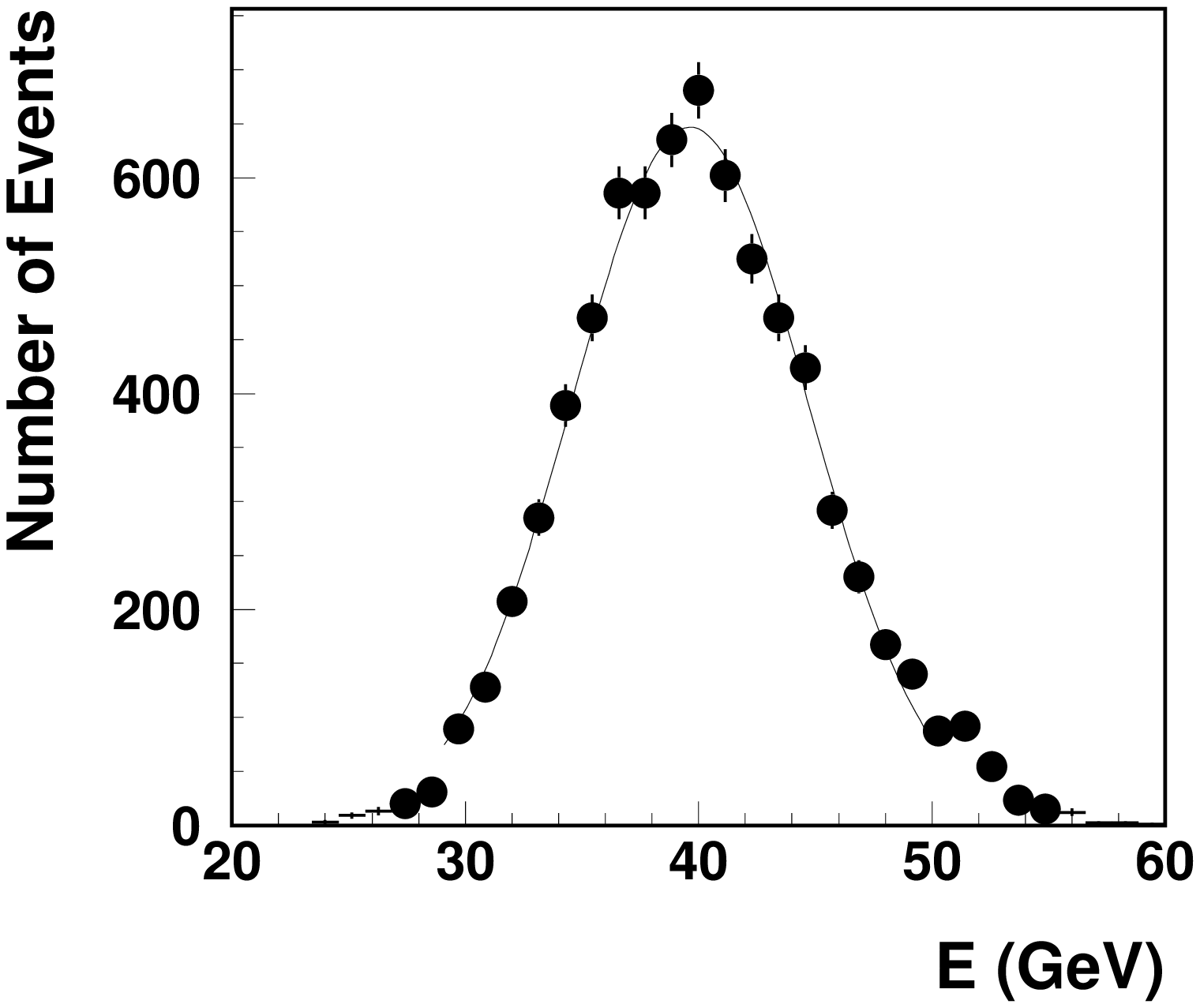,width=0.21\textwidth,height=0.5\textheight}
&
\epsfig{figure=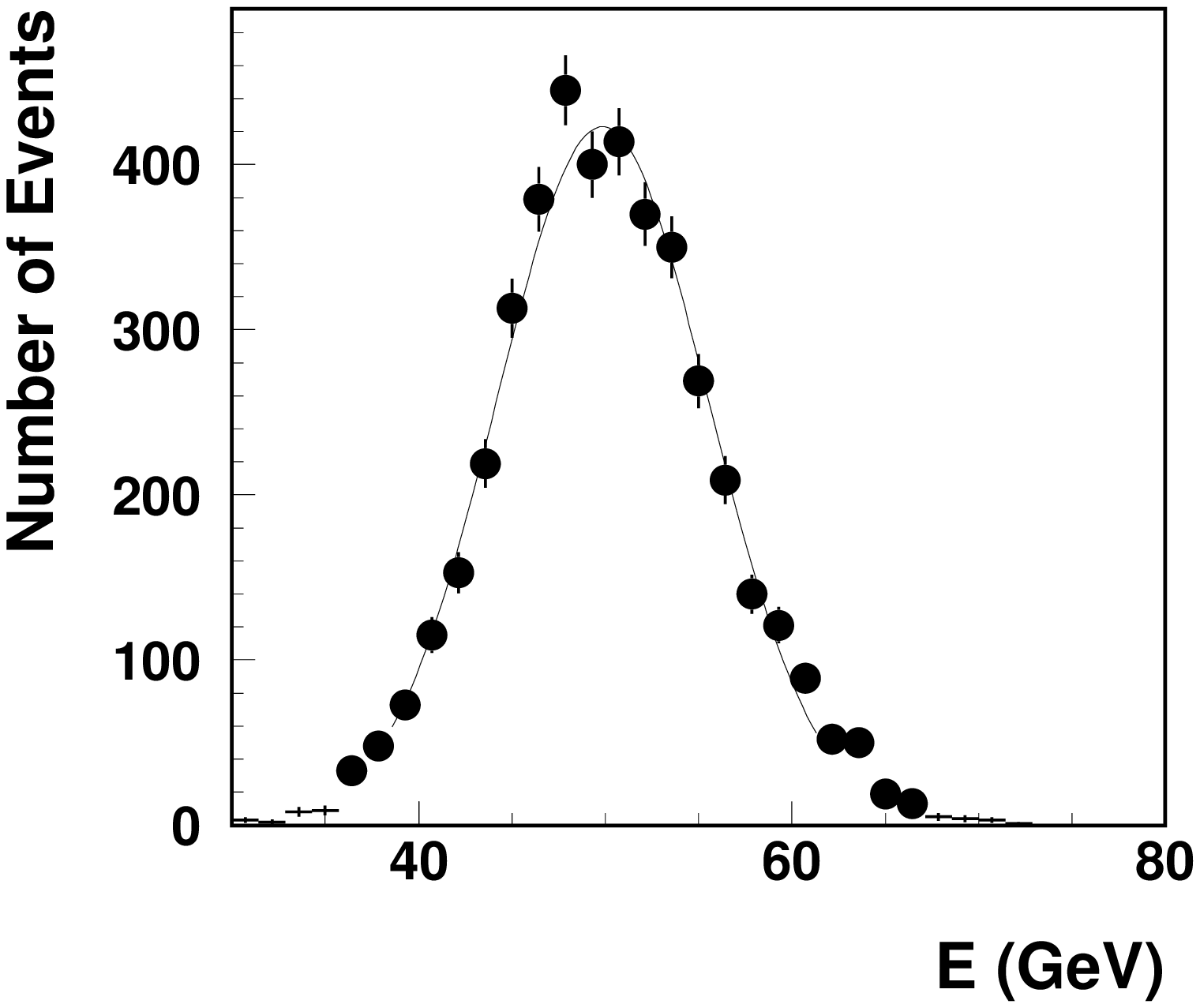,width=0.21\textwidth,height=0.5\textheight} 
\\[-15mm]
\epsfig{figure=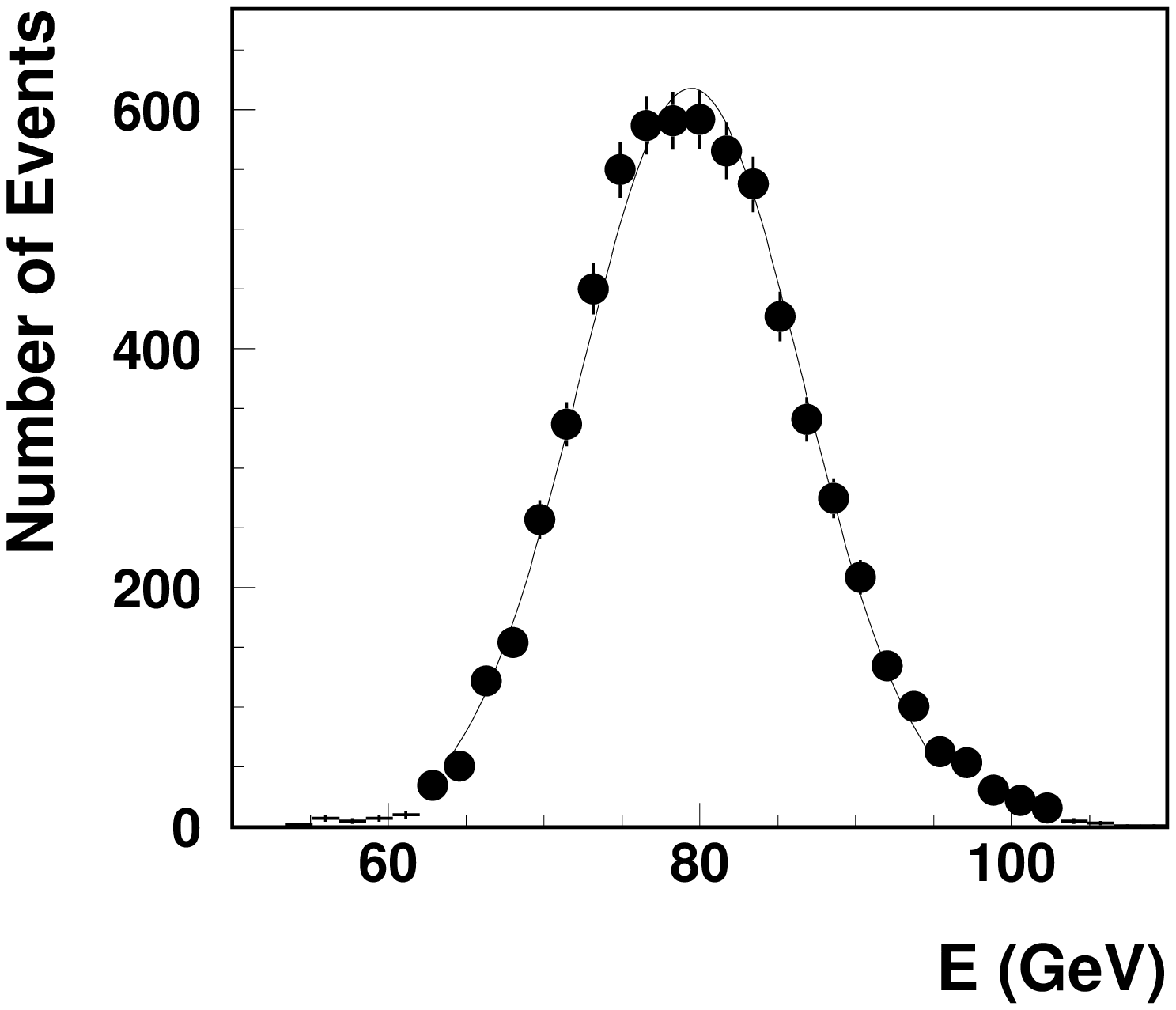,width=0.21\textwidth,height=0.5\textheight}
&
\epsfig{figure=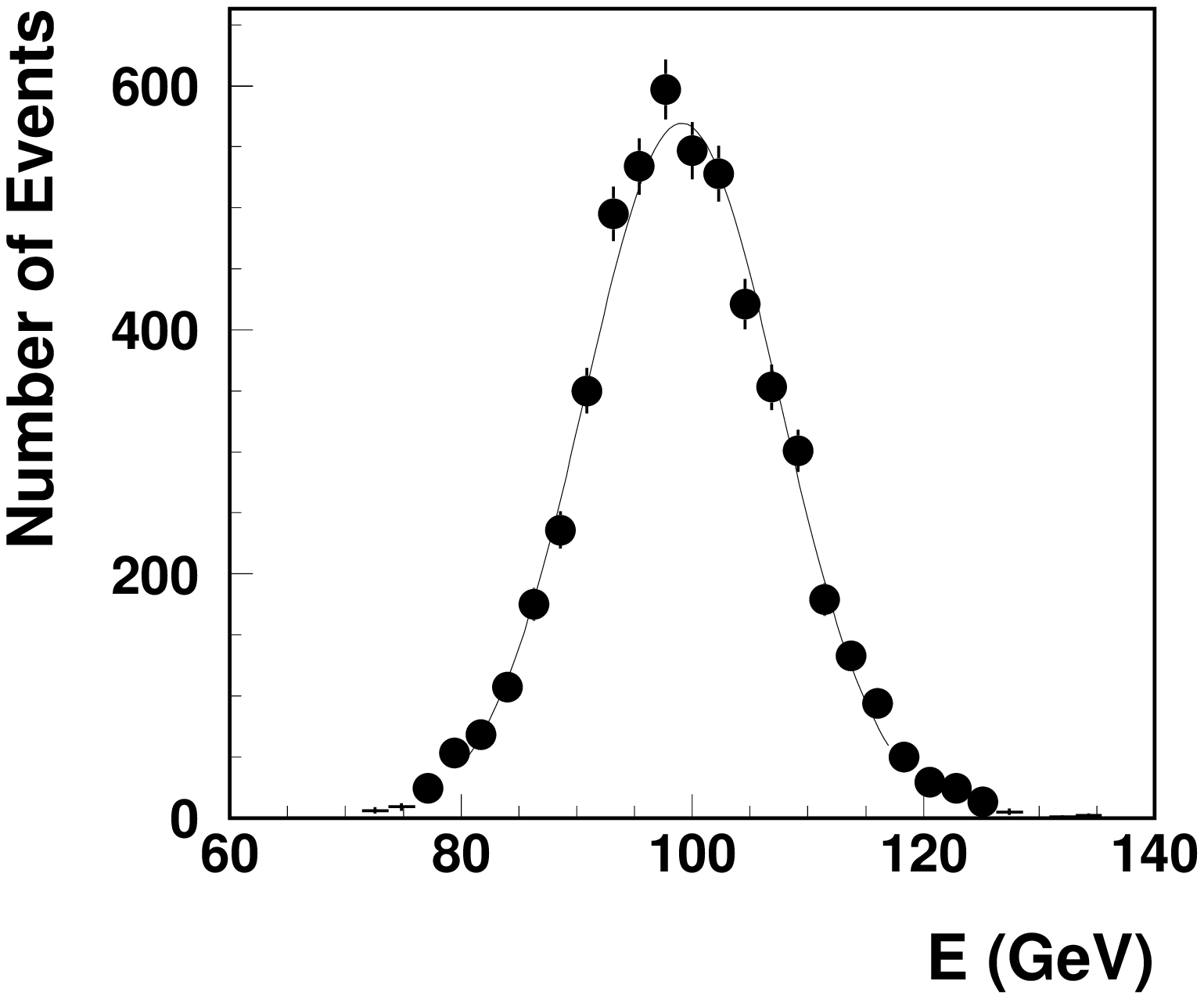,width=0.21\textwidth,height=0.5\textheight} 
&
\epsfig{figure=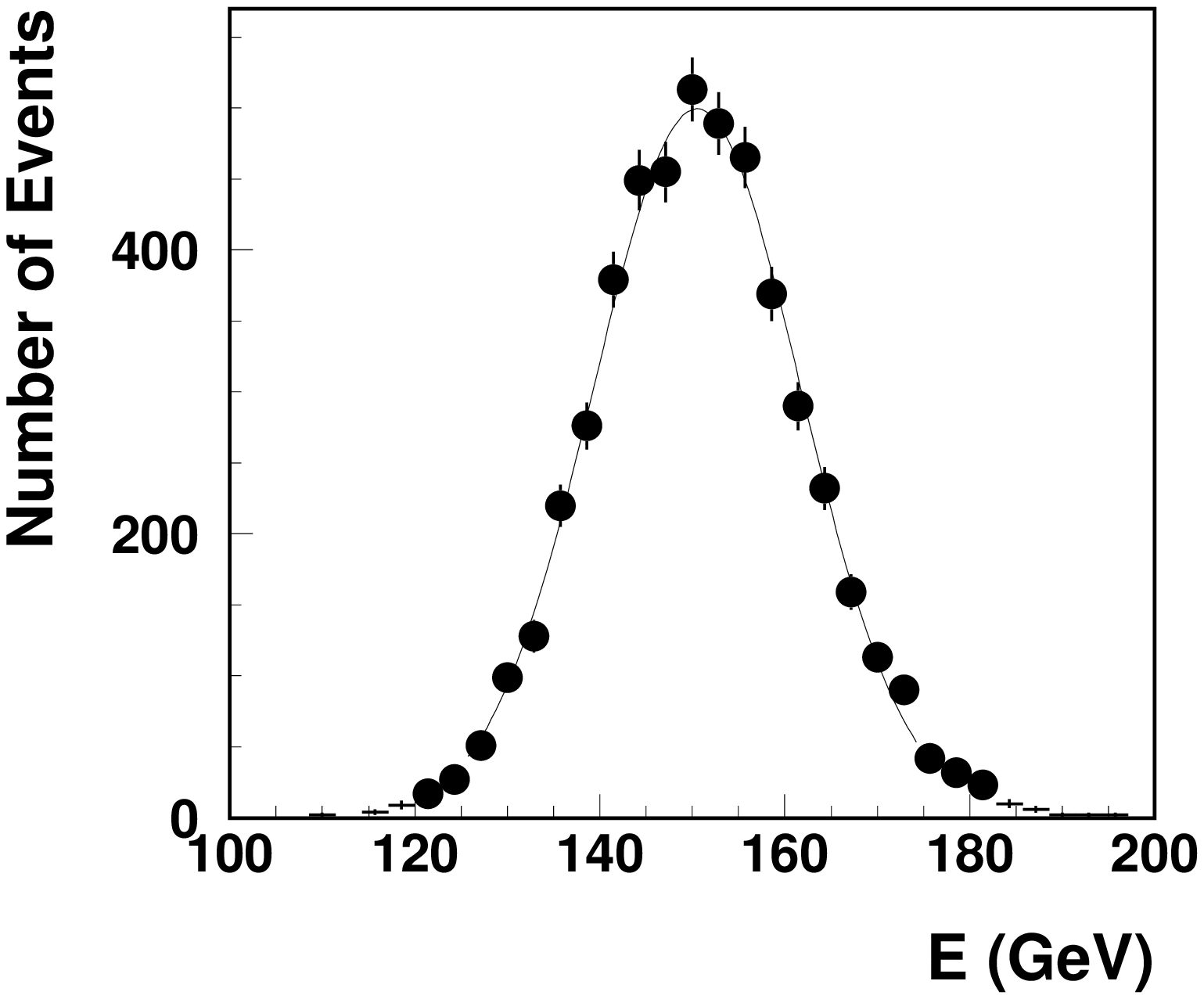,width=0.21\textwidth,height=0.5\textheight}
&
\epsfig{figure=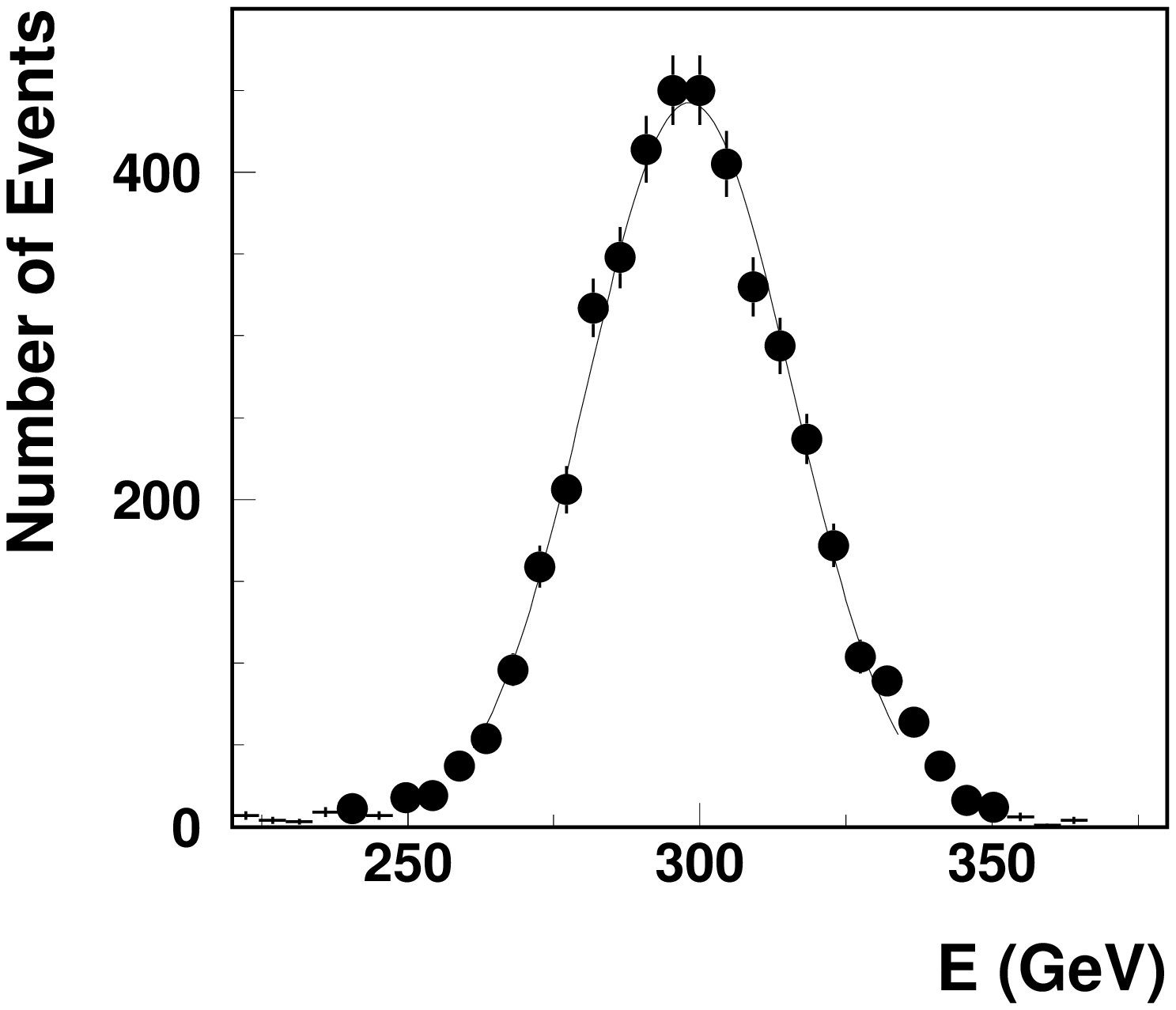,width=0.21\textwidth,height=0.5\textheight} 
\\[-10mm]
\end{tabular}
\end{center}
       \caption{
                The energy distributions  for 
                $E_{beam}$ = 10, 20, 40, 50 GeV
                (top row, left to right) 
                and $E_{beam}$ = 80, 100, 150, 300  GeV 
                (bottom row, left to right).  
       \label{f01}}
\end{figure*}

Fig.\ \ref{f01} shows the pion energy spectra reconstructed with the $e/h$ 
method ($\epsilon = 0.1\%$).
The mean and $\sigma$ values of these distributions are extracted with 
Gaussian fits over $\pm 2 \sigma$ range.
The obtained mean values $E$, the energy resolutions $\sigma$, and the 
fractional energy resolutions $\sigma / E$ are listed in Table \ref{tv1} 
for the various beam energies.

\subsection{Energy Linearity}
\begin{figure*}[tbph]
\begin{center}   
\begin{tabular}{c}
\epsfig{figure=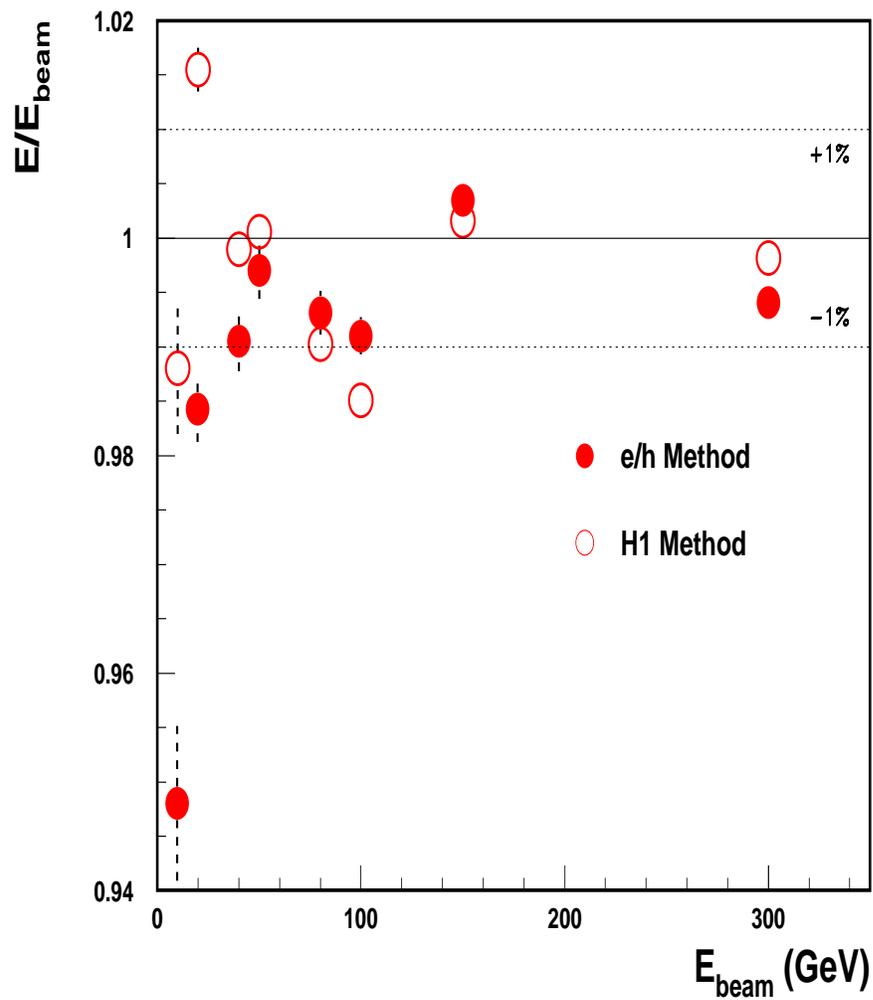,width=0.95\textwidth,height=0.9\textheight} 
\\[-20mm]
\end{tabular}
\end{center}
       \caption{
         Energy linearity as a function of the beam energy for 
         the $e/h$ method  
         (black circles) and the cells weighting H1 method (open circles).
       \label{f03}}
\end{figure*}

Fig.\ \ref{f03} demonstrates the correctness of the mean energy 
reconstruction.
The mean value of $E / E_{beam}$ is equal to $(99.5\pm0.3) \%$ and the 
spread is $\pm 1\%$ except for the point at 10 $GeV$.
But, as noted in 
\cite{comb96}, 
at this point the result is strongly dependent on the effective capability 
to remove events with interactions in the dead material upstream and to 
deconvolve the real pion contribution from the muon contamination. 
Fig.\ \ref{f03} also shows the comparison of the linearity, $E / E_{beam}$, 
as a function of the beam energy for the $e/h$ method and for the cells 
weighting method 
\cite{comb96}.
As can be seen, the comparable quality of the linearity is observed for 
these two methods.
 
\subsection{Energy Resolutions}
\begin{figure*}[tbph]
\begin{center}   
\begin{tabular}{c}
\epsfig{figure=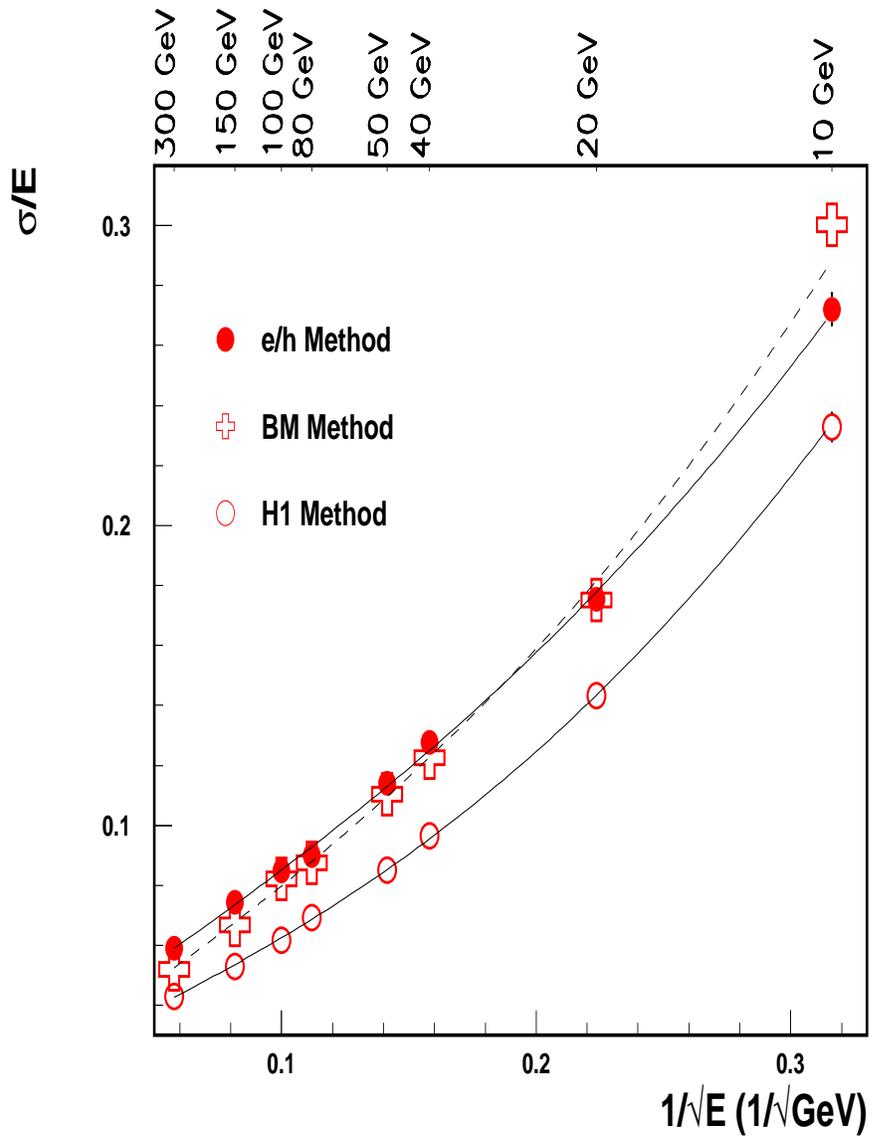,width=0.95\textwidth,height=0.9\textheight}
\\[-20mm]
\end{tabular}
\end{center}
      \caption{
      The energy resolutions obtained with the $e/h$ method (black circles),
          the benchmark method (crosses) and the cells weighting method 
         (circles).          
       \label{f05}}
\end{figure*}

Fig.\ \ref{f05} shows the fractional energy resolutions ($\sigma / E$) as a 
function of $1 / \sqrt{E_{beam}}$ obtained by three methods: the $e/h$ 
method (black circles), the benchmark method \cite{comb96} (crosses), and
the cells weighting method \cite{comb96} (open circles).
As can be seen, the energy resolutions for the $e/h$ method are comparable 
with the benchmark method and only of $30 \%$ worse than for the cells 
weighting method. 
A fit to the  data points gives the fractional energy resolution for the 
$e/h$ method obtained using the iteration procedure with 
$\epsilon = 0.1\%$:
$\sigma/E =[(58\pm3)\%\sqrt{GeV}/\sqrt{E}
    + (2.5\pm0.3)\%]\oplus (1.7\pm0.2)\ GeV/E$,       
for the $e/h$ method using the first approximation:
$\sigma/E =[(56\pm3)\%\sqrt{GeV}/\sqrt{E}
     + (2.7\pm0.3)\%]\oplus (1.8\pm0.2)\ GeV/E$,
for the benchmark method of 
$\sigma/E =[(60\pm3)\%\sqrt{GeV}/\sqrt{E}
    + (1.8\pm0.2)\%]\oplus (2.0\pm0.1)\ GeV/E$,
for the cells weighting  method of 
$\sigma/E =[(42\pm2)\%\sqrt{GeV}/\sqrt{E}
     + (1.8\pm0.1)\%]\oplus (1.8\pm0.1)\ GeV/E$,       
where the symbol $\oplus$ indicates a sum in quadrature.
As can be seen, the sampling term is consistent within errors for the $e/h$ 
method and the benchmark method and is smaller by 1.5 times for the cells 
weighting method.
The constant term is the same for the benchmark method and the cells 
weighting method and is larger by $(0.7\pm0.3) \%$ for the $e/h$ method.
The noise term of about $1.8\ GeV$ is the same for these three methods that 
reflects its origin as the electronic noise.
As to the two approaches for the $e/h$ method, the fitted parameters 
coincide within errors.

\section{Hadronic Shower Development}
\begin{figure*}[tbph]    
\begin{center}
\begin{tabular}{c}
\mbox{\epsfig{figure=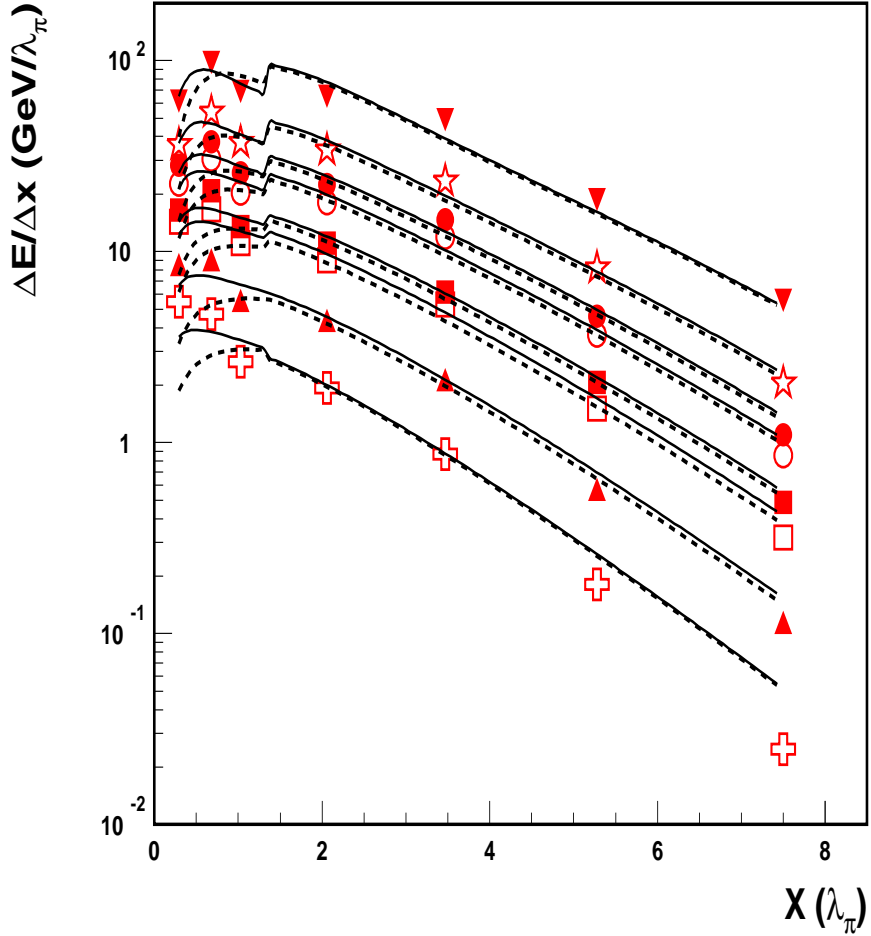,width=0.95\textwidth,height=0.85\textheight}}
\\[-20mm]
\end{tabular}
\caption{
        The experimental differential longitudinal energy depositions at
        10 GeV (crosses), 20 GeV (black top triangles), 
        40 GeV (open squares),
     50 GeV (black squares), 80 GeV (open circles), 100 GeV (black circles),
        150 GeV (stars), 300 GeV (black bottom triangles)
        energies as a function of the longitudinal coordinate $x$ in units
        $\lambda_{\pi}$ for the combined calorimeter and the results 
        of the description by the Bock(dashed lines) and modified 
        (solid lines) parameterizations.
        }
\label{fv6-1}
\end{center}
\end{figure*}

We used this energy reconstruction method and obtained the energy 
depositions, $E_i$, in each longitudinal sampling with the thickness of 
$\Delta x_i$ in units $\lambda_{\pi}$.
Table \ref{T1} lists and Fig.\ \ref{fv6-1} shows the differential energy 
depositions $(\Delta E/ \Delta x)_i = E_i / \Delta x_i$ as a function of 
the longitudinal coordinate $x$ for 10 -- 300 GeV.

\subsection{Longitudinal Hadronic Shower Parameterization}
There is the well known parameterization of the longitudinal had\-ro\-nic 
shower development from the shower origin suggested in
\cite{bock81} 
\begin{equation}
        \frac{dE_{s} (x)}{dx} =
                N\
                \Biggl\{
                w\ \biggl( \frac{x}{X_0} \biggr)^{a-1}\
                e^{- b \frac{x}{X_0}}\
                + \
                (1-w)\
                \biggl( \frac{x}{\lambda_I} \biggr)^{a-1}\
                e^{- d \frac{x}{\lambda_I}}
                \Biggr\} \ ,
\label{elong00}
\end{equation}
where $X_0$ is the radiation length, $\lambda_I$ is the interaction length,
$N$ is the normalization factor, $a,\ b,\ d,\ w$ are parameters:
$a = 0.6165 + 0.3193\ lnE$, $b = 0.2198$, $d = 0.9099 - 0.0237\ lnE$,
$\omega = 0.4634$.

This parameterization is from the shower origin.
But our data are from the calorimeter face and due to the unsufficient 
longitudinal segmentation can not be transformed to the shower origin.
Therefore, we used the analytical representation of the hadronic shower
longitudinal development from the calorimeter face 
\cite{kulchitsky98}:
\begin{eqnarray}
        \frac{dE (x)}{dx} & = &
                N\
                \Biggl\{
                \frac{w X_0}{a}
                \biggl( \frac{x}{X_0} \biggr)^a
                e^{- b \frac{x}{X_0}}
                {}_1F_1 \biggl(1,a+1,
                \biggl(b - \frac{X_0}{\lambda_I} \biggr) \frac{x}{X_0}
                \biggr)
                \nonumber \\
                & & + \
                \frac{(1 - w) \lambda_I}{a}
                \biggl( \frac{x}{\lambda_I} \biggr)^a
                e^{- d \frac{x}{\lambda_I}}
                {}_1F_1 \biggl(1,a+1,
                \bigl( d -1 \bigr) \frac{x}{\lambda_I} \biggr)
                \Biggr\} ,
\label{elong03}
\end{eqnarray}
here ${}_1F_1(\alpha,\beta,z)$ is the confluent hypergeometric function.

Note that the formula (\ref{elong03}) is given for a calorimeter 
characterizing by the certain $X_0$ and $\lambda_I$ values.
At the same time, the values of $X_0$, $\lambda_I$ and the $e/h$ ratios
are different for electromagnetic and hadronic compartments of a combined 
calorimeter. 
So, it is impossible straightforward use of the formula (\ref{elong03}) for 
the description of a hadronic shower longitudinal profiles in combined 
calorimetry.

In \cite{kulchitsky99} suggested the following algorithm of combination of 
the electromagnetic calorimeter ($em$) and hadronic calorimeter ($had$) 
curves of the differential longitudinal energy deposition $dE/dx$. 
At first, a hadronic shower develops in the electromagnetic calorimeter to 
the boundary value $x_{em}$ which corresponds to certain integrated 
measured energy $E_{em}(x_{em})$.
Then, using the corresponding integrated hadronic curve, $E(x) =$ 
\linebreak[4]
$\int_0^x (dE/dx) dx$, the point $x_{had}$ is found from equation 
$E_{had}(x_{had}) = E_{em}(x_{em})$ $+ E_{dm}$.
From this point a shower continues to develop in the hadronic calorimeter.
In principle, instead of the measured value of $E_{em}$ one can use the 
calculated value of $E_{em} = \int_0^{x_{em}} (dE/dx) dx$ obtained from the 
integrated electromagnetic curve. 
In this way, the combined curves have been obtained. 

Fig.\ \ref{fv6-1} shows the differential energy depositions
$(\Delta E/ \Delta x)_i = E_i / \Delta x_i$ as a function of the 
longitudinal coordinate $x$ in units $\lambda_{\pi}$ for the 10 -- 300 GeV 
and comparison with the combined curves for the longitudinal hadronic 
shower profiles (the dashed lines).
It can be seen that there is a significant disagreement between the 
experimental data and the combined curves in the region of the LAr 
calorimeter and especially at low energies.

\subsection{Modification of Shower Parameterization}
We attempted to improve the description and to include such essential 
feature of a calorimeter as the $e/h$ ratio.
Several modifications and adjustments of some parameters of this 
parameterization have been tried.
It turned  out that the changes of two parameters $b$ and $w$ in the formula
(\ref{elong03}) in such a way that 
$b = 0.22 \cdot (e/h)_{cal} / (e/h)_{cal}^{\prime}$ and 
$w = 0.6  \cdot (e/\pi)_{cal} / (e/\pi)_{cal}^{\prime}$
made it possible to obtain the reasonable description of the experimental 
data.
Here the values of the $(e/h)_{cal}^{\prime}$ ratios are 
$(e/h)^{\prime}_{em} \approx 1.1$ and $(e/h)^{\prime}_{had} \approx 1.3$
which correspond to the data used for the Bock et al.\ parameterization 
\cite{bock81}.
The $(e/\pi)_{cal}^{\prime}$ are calculated using formula (\ref{ev10}). 

In Fig.\ \ref{fv6-1} the experimental differential longitudinal energy 
depositions and the results of the description by the modified 
parameterization (the solid lines) are compared.
There is a reasonable agreement (probability of description is more than 
$5\%$) between the experimental data and the curves taking into account 
uncertainties in the parametrization function \cite{bock81}.
In such case the Bock et al.\ parameterization is the private case for some 
fixed the $e/h$ ratio.

\subsection{Energy Deposition in Compartments}
\begin{figure*}[tbph]
\begin{center}   
\begin{tabular}{c}
\epsfig{figure=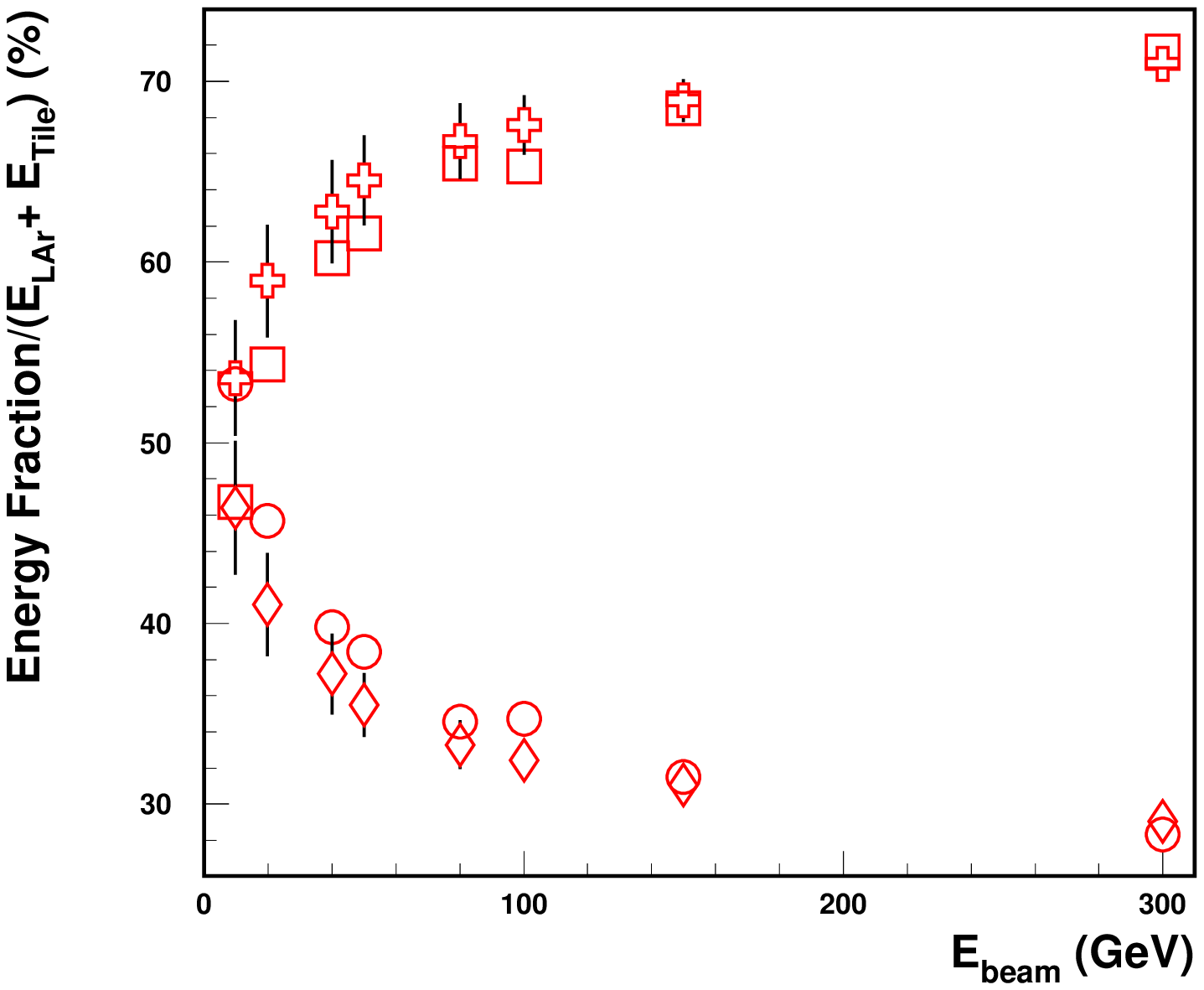,width=0.95\textwidth,height=0.9\textheight} 
\\[-20mm]
\end{tabular}
\end{center}
       \caption{ 
         Energy deposition (percentage) in the LAr and Tile 
         calorimeters at different beam energies.
         The circles (squares) are the measured energy depositions
         in the LAr (Tile) calorimeter, the diamonds (crosses) are the 
         calculated energy depositions in the ones.
        }
       \label{f04-a}
\end{figure*}

The obtained parameterization has some additional applications.
For example, this formula may be used for an estimate of the energy 
deposition in various parts of a combined calorimeter.
This demonstrates in Fig.\ \ref{f04-a} in which the measured and calculated 
relative values of the energy deposition in the LAr and Tile calorimeters 
are presented.
The relative energy deposition in the LAr calorimeter decreases from about 
50\% at 10 $GeV$ to 30\% at 300 $GeV$.
On the contrary, the one in Tile calorimeter increases with the energy
increasing.

\section{Conclusions}
Hadron energy reconstruction for the ATLAS barrel prototype combined 
calorimeter, consisting of the lead-liquid argon electromagnetic part and 
the iron-scintillator hadronic part, in the framework of the 
non-pa\-ra\-met\-ri\-cal method has been fulfilled.
The non-parametrical method of the energy reconstruction for a combined 
calorimeter uses only the known $e/h$ ratios and the electron calibration 
constants, does not require the determination of any parameters by a 
minimization technique and can be used for the fast energy reconstruction 
in the first level trigger. 
The correctness of the reconstruction of the mean values of energies 
(for energy biger than 10 GeV) within $\pm 1\%$ has been demonstrated.
The obtained fractional energy resolution is 
$[(58\pm3)\%\sqrt{GeV}/\sqrt{E}+(2.5\pm0.3)\%]\oplus (1.7\pm0.2)\ GeV/E$.
The obtained value of the $e/h$ ratio for electromagnetic compartment of 
the combined calorimeter is $1.74\pm0.04$ and agrees with the 
prediction that $e/h > 1.7$ for this electromagnetic calorimeter.
The results of the study of the longitudinal hadronic shower development
are presented.
The data have been taken on the H8 beam of the CERN SPS, with the pion 
beams of 10 -- 300 GeV.

\section{Acknowledgments}
This work is the result of the efforts of many people from the ATLAS
Collaboration.
The authors are greatly indebted to all Collaboration for their test beam 
setup and data taking.
Authors are grateful Peter Jenni for fruitful discussion. 



\begin{table}[tbh]
\begin{center}
\caption{
        Mean reconstructed energy, energy resolution and 
        fractional energy resolution for the various beam energies.
        }
\label{tv1}
\begin{tabular}{|r|c|c|c|}
\multicolumn{4}{c}{\mbox{~~}}\\[-3mm]
\hline
$E_{beam}$&$E$ (GeV)&$\sigma$ (GeV)&$\sigma / E\ (\%)$\\ 
\hline
$10^{\ast}$     GeV&  $9.30\pm0.07$& $2.53\pm0.05$&$27.20\pm0.58$\\
$20^{\star}$    GeV& $19.44\pm0.06$& $3.41\pm0.06$&$17.54\pm0.31$\\
40              GeV& $39.62\pm0.11$& $5.06\pm0.08$&$12.77\pm0.21$\\
50              GeV& $49.85\pm0.13$& $5.69\pm0.13$&$11.41\pm0.26$\\
80              GeV& $79.45\pm0.16$& $7.14\pm0.14$& $8.99\pm0.18$\\
100             GeV& $99.10\pm0.17$& $8.40\pm0.16$& $8.48\pm0.16$\\
150             GeV&$150.52\pm0.19$&$11.20\pm0.18$& $7.44\pm0.12$\\
300             GeV&$298.23\pm0.37$&$17.59\pm0.33$& $5.90\pm0.11$\\
\hline
\multicolumn{4}{@{}l@{}}{$^{\ast}$The measured value of the 
                                               beam energy is 9.81 $GeV$.}\\
\multicolumn{4}{@{}l@{}}{$^{\star}$The measured value of the 
                                        beam energy is 19.8 $GeV$.}     \\
\end{tabular}
\end{center}
\end{table}
\begin{table}[tbh]
\vspace*{-10mm}
\begin{center}
\caption{
        The differential energy depositions $\Delta E/ \Delta x$
        as a function of the longitudinal coordinate $x$ for the various 
        beam energies.
        } 
\label{T1}
\begin{tabular}{|c|c|c|c|c|c|}  
\multicolumn{6}{c}{\mbox{~~}}\\[-3mm]
\hline
    N&  x&\multicolumn{4}{|c|}{$E_{beam}$ (GeV)}\\
\cline{3-6}
depth&($\lambda_{\pi}$)&10  &20           &40       &50                 \\ 
\hline
1  &0.294&$5.45\pm0.08$  & $8.58\pm0.16$ &$14.3\pm0.2$   &$16.6\pm0.4$ \\
2  &0.681&$4.70\pm0.08$  & $9.10\pm0.15$ &$16.7\pm0.2$   &$20.8\pm0.3$ \\
3  &1.026&$2.66\pm0.06$  & $5.55\pm0.11$ &$11.1\pm0.2$   &$13.6\pm0.2$ \\
dm &1.315&$1.35\pm0.07$  & $2.75\pm0.14$ &$5.28\pm0.26$  &$6.46\pm0.32$ \\
4  &2.06 &$1.93\pm0.03$  & $4.35\pm0.06$ &$8.99\pm0.08$  &$11.0\pm0.1$ \\
5  &3.47 &$0.87\pm0.02$  & $2.13\pm0.04$ &$5.29\pm0.06$  &$6.15\pm0.10$ \\
6  &5.28 &$0.18\pm0.01$  & $0.57\pm0.02$ &$1.50\pm0.03$  &$2.07\pm0.05$ \\
7  &7.50 &$0.025\pm0.003$& $0.11\pm0.01$ &$0.32\pm0.01$  &$0.49\pm0.02$\\
\hline
    N&  x&\multicolumn{4}{|c|@{}}{$E_{beam}$ (GeV)}\\
\cline{3-6}
depth&($\lambda_{\pi}$)&80&100&150&300 \\ 
\hline
1  &0.294&$22.6\pm0.6$ &$28.4\pm0.6$ &$36.3\pm0.7$ &$61.3\pm1.5$        \\
2  &0.681&$30.4\pm0.4$ &$37.6\pm0.5$ &$53.5\pm0.8$ &$97.9\pm1.7$        \\ 
3  &1.026&$20.3\pm0.3$ &$25.7\pm0.4$ &$37.2\pm0.6$ &$68.9\pm1.2$        \\ 
dm &1.315&$10.1\pm0.5$ &$12.8\pm0.6$ &$19.0\pm1.0$ &$34.1\pm1.7$        \\  
4  &2.06 &$18.0\pm0.1$ &$22.4\pm0.2$ &$33.9\pm0.3$ &$64.8\pm0.7$        \\  
5  &3.47 &$11.9\pm0.1$ &$14.6\pm0.2$ &$23.3\pm0.2$ &$49.0\pm0.5$        \\
6  &5.28 &$3.66\pm0.06$&$4.57\pm0.08$&$8.18\pm0.13$&$18.6\pm0.3$        \\
7  &7.50 &$0.86\pm0.03$&$1.10\pm0.04$&$2.04\pm0.06$&$5.54\pm0.15$       \\  
\hline
\end{tabular}
\end{center}
\end{table}

\newpage
\clearpage
\clearpage
\clearpage
\clearpage
\clearpage
\clearpage
\clearpage
\clearpage
\clearpage


\begin{thebibliography}{00}
\bibitem{atcol94}
ATLAS Collaboration, ATLAS Technical Proposal for a General Purpose pp 
Experiment at the Large Hadron Collider, 
CERN/LHCC/94-93, CERN,  Geneva,  Switzerland.
\bibitem{TILECAL96}
ATLAS Collaboration, ATLAS TILE Calorimeter Technical Design Report, 
CERN/LHCC/96-42,  ATLAS TDR 3,  1996, CERN,   Geneva.
\bibitem{LARG96}
ATLAS Collaboration, ATLAS Liquid Argon Calorimeter Technical Design Report,
CERN/LHCC/96-41,  ATLAS TDR 2,  1996, CERN,   Geneva,   Switzerland.
\bibitem{ariztizabal94}
F.~Ariztizabal et al., NIM {\bf A349} (1994) 384.
\bibitem{juste95}
A.~Juste, ATL-TILECAL-95-69,  1995,  CERN,  Geneva,  Switzerland.
\bibitem{budagov96-72}
J.~Budagov,   Y.~Kulchitsky et al., 
ATL-TILECAL-96-72,  1996,  CERN,  Geneva,  Switzerland.
\bibitem{kulchitsky99-12}
Y.~Kulchitsky et al., 
ATL-TILECAL-99-002,  1999,  CERN,  Geneva.
\bibitem{comb96}
ATLAS Collaboration, 
{\it Results from an Expanded Combined Test of the Electromagnetic Liquid
Argon Calorimeter with a Hadronic Scintillating-Tile Calorimeter},
Submitted to NIM {\bf A}, 2000.
\bibitem{cobal98}
M.~Cobal et al.,
ATL-TILECAL-98-168, 1998, CERN, Geneva.
\bibitem{ccARGON} 
D.M.~Gingrich et al., (RD3 Collaboration), NIM {\bf A364} (1995) 290. 
\bibitem{gild91} 
O.~Gildemeister, F.~Nessi-Tedaldi, M.~Nessi,
Proc.\ 2nd Int.\ Conf.\ on Calorimetry in High Energy Physics, Capri, 1991.
\bibitem{ccNIM} 
F.~Ariztizabal et al., RD34 Collaboration, NIM (1994) 384;\\
E.Berger et al., RD34 Collaboration, CERN/LHCC/95-44.
\bibitem{ccrd34rep94} 
M.~Bosman et al., RD34 Collaboration, CERN/DRDC/93-3, 1993; \\
F.Ariztizabal et al., RD34 Collaboration, CERN/DRDC/94-66, 1994.
\bibitem{shower98}
S.~Agnvall et al.,  
{\it Hadronic Shower Development in Iron-Scintillator Tile Calorimetry},
NIM {\bf A} (in press).
\bibitem{budagov-97-127}
J.~Budagov,  Y.~Kulchitsky et al.,
ATL-TILECAL-97-127, 1997, CERN, Geneva, Switzerland.
\bibitem{wigmans88}
R.~Wigmans,  NIM {\bf A265} (1988) 273.
\bibitem{groom89}
D.~Groom, Proceedings of the Workshop on Calorimetry for the Supercollides, 
Tuscaloosa, Alabama, USA, 1989.
\bibitem{combined94}
Z.~Ajaltouni et al.,  NIM {\bf A387} (1997) 333.
\bibitem{bosman99}
M.~Bosman,  Y.~Kulchitsky,  M.~Nessi, 
ATL-COM-TILECAL-99-011, 1999, CERN, Geneva, Switzerland.
\bibitem{atcol99}
ATLAS Collaboration, ATLAS Physical Technical Design Report, v.1,  
CERN-LHCC-99-02,  ATLAS-TDR-14, CERN,   Geneva,   Switzerland
\bibitem{wigmans91}
R.Wigmans, 
Proc.2nd Int.Conf.\ on Calorimetry in HEP, Capri,  1991.
\bibitem{bock81} 
R.~Bock et al., NIM {\bf 186} (1981) 533.
\bibitem{kulchitsky98}
Y.~Kulchitsky et al., NIM {\bf A413} (1998) 484.
\bibitem{kulchitsky99}
Y.~Kulchitsky et al., Submitted to NIM {\bf A}, 2000.
\end{thebibliography}
\end{document}